\documentclass[10pt,journal,compsoc]{IEEEtran}

\ifCLASSOPTIONcompsoc
  \usepackage[nocompress]{cite}
\else
  \usepackage{cite}
\fi

\usepackage{blindtext}
\usepackage{rotating} 

\usepackage{hyperref}   

\usepackage{multirow}
\usepackage[table,xcdraw]{xcolor}
\usepackage[utf8]{inputenc} 
\usepackage[T1]{fontenc}    
\usepackage{url}            
\usepackage{ragged2e}
\usepackage{tabularx}
\usepackage{booktabs}       
\usepackage{amsfonts}       
\usepackage{nicefrac}       
\usepackage{microtype}      
\usepackage{lipsum}
\usepackage{multicol}
\usepackage{soul}
\usepackage{color}
\usepackage{wallpaper}
\usepackage{multirow}
\usepackage{graphicx}
\usepackage{rotating}
\usepackage{array,makecell}
\usepackage{tablefootnote}
\usepackage[nohyperlinks,nolist]{acronym}
\usepackage{enumitem}
\usepackage{amsmath}
\usepackage{cleveref}

\usepackage{pifont}
\newcommand{\cmark}{\ding{52}}%
\newcommand{\xmark}{\ding{56}}%

\setlist{nolistsep}

\definecolor{orangeX}{rgb}{1,.5,0}
\definecolor{blueX}{rgb}{.2, .59, .88}
\definecolor{purpleX}{rgb}{.294118, 0, .509804}
\definecolor{greenX}{rgb}{.721, .878, .341}
\definecolor{bole}{rgb}{0.47, 0.27, 0.23}
\definecolor{mypink3}{cmyk}{0, 0.7808, 0.4429, 0.1412}

\newcommand{\eg}{e.\,g.\,, }
\newcommand{\ie}{i.\,e.\,, }

\newcommand{\cf}{{cf.\,}}
\newcommand{\musec}{MuSe-CaR\,}
\newcommand{\musewild}{MuSe-Wild\,}
\newcommand{\musetopic}{MuSe-Topic\,}
\newcommand{\musetrust}{MuSe-Trust\,}

\newcommand{\egm}{\textsc{eGeMAPS\,}}
\newcommand{\vgg}{\textsc{VGGish\,}}
\newcommand{\vggf}{\textsc{VGGFace\,}}
\newcommand{\ft}{\textsc{FastText\,}}
\newcommand{\bert}{\textsc{BERT\,}}
\newcommand{\au}{\textsc{FAU\,}}
\newcommand{\mtcnn}{\textsc{MTCNN\,}}

\usepackage[symbol]{footmisc}

\usepackage{tcolorbox}
\tcbuselibrary{skins}
\makeatletter
\def\blfootnote{\gdef\@thefnmark{}\@footnotetext}
\makeatother
\newcommand{\hand}[1] {
    \begin{tcolorbox}[skin=enhanced,
                    width=2in,
                    colback=white,
                    fontlower=\sffamily,
                    fontupper=~\rmfamily,
                    middle=0mm,
                    center
                    ]
    \includegraphics[width=0.46cm]{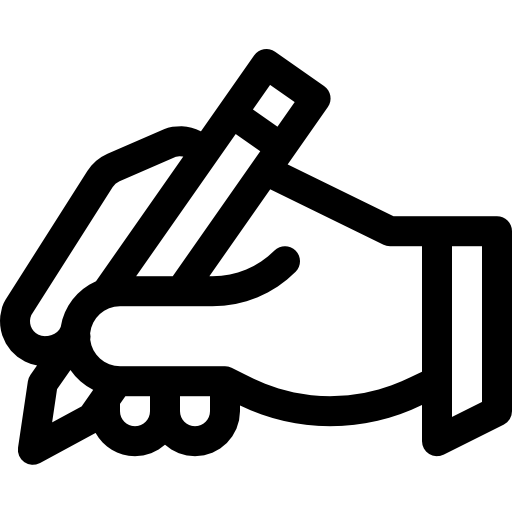}
    #1
    \end{tcolorbox}
}
\newcommand{\google}[1] {
    \begin{tcolorbox}[skin=enhanced,
                    width=2in,
                    colback=white,
                    fontlower=\sffamily,
                    fontupper=~\rmfamily,
                    middle=0mm,
                    center
                    ]
    \includegraphics[width=0.51cm,trim={.1cm .5cm 1cm .5cm},clip]{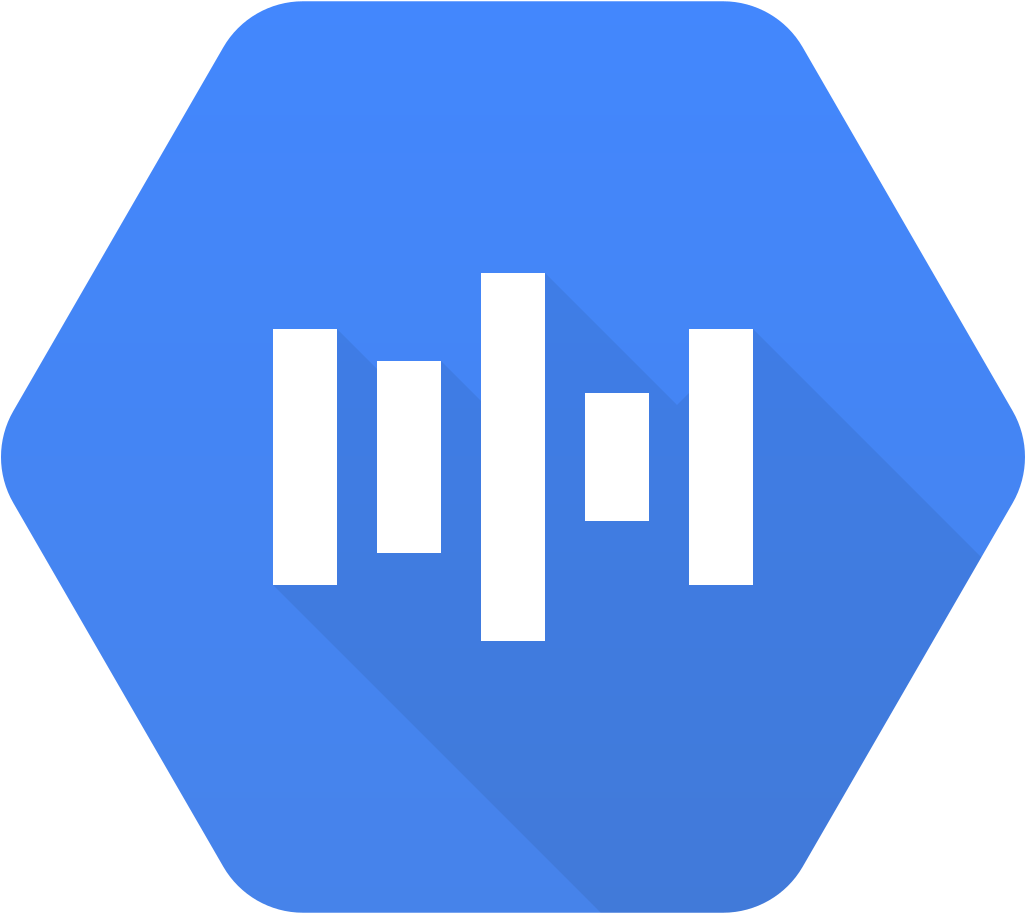}
    #1
    \end{tcolorbox}
}
\newcommand{\aws}[1] {
    \begin{tcolorbox}[skin=enhanced,
                    width=2in,
                    colback=white,
                    fontlower=\sffamily,
                    fontupper=~\rmfamily,
                    middle=0mm,
                    center
                    ]
    \includegraphics[width=0.45cm,trim={.1cm 1cm 9.5cm 1.3cm},clip]{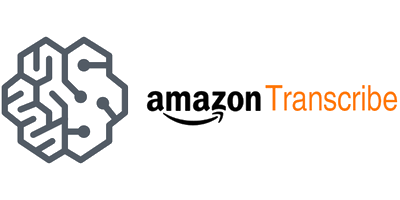}
    #1
    \end{tcolorbox}
}
\begin{document}

\begin{acronym}
\acro{AFEW}[AFEW]{Acted Facial Expressions In The Wild}
\acro{AUC}[AUC]{area under the curve}
\acro{CNN}[CNN]{Con\-vo\-lu\-tion\-al Neural Network}
\acrodefplural{CNN}[CNNs]{Con\-vo\-lu\-tion\-al Neural Networks}
\acro{DCTW}[DCTW]{Deep Canonical Time Warping}
\acro{DNN}[DNN]{Deep Neural Network}
\acro{DTW}[DTW]{Dynamic Time Warping}
\acro{EER}[EER]{equal error rate}
\acro{EWE}[EWE]{Evaluator Weighted Estimator}
\acro{FAU}[FAU]{Facial Action Unit}
\acrodefplural{FAU}[FAUs]{Facial Action Units}
\acro{GMM}[GMM]{Gaussian mixture model}
\acro{HCI}[HCI]{human computer interaction}
\acro{HMDB}[HMDB]{Human Motion Database}
\acro{LeakyReLU}[LeakyReLU]{Leaky Rectified Linear Unit}
\acro{LSTM}[LSTM]{Long Short-Term Memory Recurrent Neural Network}
\acro{MLP}[MLP]{Multilayer Perceptron}
\acro{MOSI}[MOSI]{Multimodal Corpus of Sentiment Intensity}
\acro{MuSe-CaR}[MuSe-CaR]{multimodal sentiment analysis in Car Reviews}
\acro{ReLU}[ReLU]{rectified linear unit}
\acro{RMSE}[RMSE]{root mean square error}
\acro{RNN}[RNN]{Recurrent Neural Network}
\acro{S2SAE}[S2SAE]{Sequence to Sequence Autoencoder}
\acro{SSD}[SSD]{Single Shot Detector}
\acro{UAR}[UAR]{unweighted average recall}
\end{acronym}

\title{The Multimodal Sentiment Analysis \\in Car Reviews (MuSe-CaR) Dataset: \\Collection, Insights and Improvements}



\author{Lukas Stappen\textsuperscript{\dag},~\IEEEmembership{Member,~IEEE,}
        Alice Baird\textsuperscript{\dag},~\IEEEmembership{Member,~IEEE,}
        Lea Schumann, 
        and~\\Bj{\"o}rn Schuller,~\IEEEmembership{Fellow,~IEEE}
\IEEEcompsocitemizethanks{\IEEEcompsocthanksitem All authors are with the EIHW -- Chair of Embedded Intelligence for Health Care and Wellbeing, University of Augsburg, Germany.\protect\\
Contact E-mail: {stappen,baird,schuller}@ieee.org
\IEEEcompsocthanksitem Bj{\"o}rn Schuller is also with GLAM -- Group on Language, Audio, \& Music, Imperial College, London, UK.}
\thanks{\dag These authors equally contributed to the design of the dataset.}}

\markboth{Journal of IEEE Transactions on Affective Computing (Early Access)}
{Shell \MakeLowercase{\textit{et al.}}: Bare Advanced Demo of IEEEtran.cls for IEEE Computer Society Journals}
%



\IEEEtitleabstractindextext{%
\begin{abstract}
Truly real-life data presents a strong, but exciting challenge for sentiment and emotion research. The high variety of possible `in-the-wild' properties makes large datasets such as these indispensable with respect to building robust machine learning models. A sufficient quantity of data covering a deep variety in the challenges of each modality to force the exploratory analysis of the interplay of all modalities has not yet been made available in this context. 
In this contribution, we present MuSe-CaR, a first of its kind multimodal dataset. The data is publicly available as it recently served as the testing bed for the 1st Multimodal Sentiment Analysis Challenge, and focused on the tasks of emotion, emotion-target engagement, and trustworthiness recognition by means of comprehensively integrating the audio-visual and language modalities. 
Furthermore, we give a thorough overview of the dataset in terms of collection and annotation, including annotation tiers not used in this year's MuSe 2020. In addition, for one of the sub-challenges --  predicting the level of trustworthiness -- no participant outperformed the baseline model, and so we propose a simple, but highly efficient Multi-Head-Attention network that exceeds using multimodal fusion the baseline by around 0.2 CCC (almost 50 \% improvement). 

\end{abstract}

}

\maketitle

\IEEEdisplaynontitleabstractindextext

%
\IEEEpeerreviewmaketitle

\ifCLASSOPTIONcompsoc
\IEEEraisesectionheading{\section{Introduction}\label{sec:introduction}}
\else
\section{Introduction}
\label{sec:introduction}
\fi

Global video traffic is estimated to grow four-fold in the coming years~\cite{cisco}, accounting for 80\,\% of all online traffic in 2019~\cite{roesler}. On social media, users view eight billion videos daily on Facebook~\cite{techcrunch} and YouTube has become the second biggest social network with nearly two billion active users and one billion hours watched each day~\cite{rank}. The internet has undergone a rapid transformation from a largely text-based Web 2.0 to a multimedia, user content-driven net. However, extracting, processing, and analysing relevant information from the huge amounts of semi-structured user-generated data available remains a challenge~\cite{bhardwaj2016text}.

Text-based sentiment analysis is now widely used, \eg for brand perception or customer satisfaction assessment, as machine learning approaches are able to learn rich text representations from data that can be applied to sentiment classification~\cite{mikolov2013distributed, camacho2017role}. However, the increased availability of other modalities (\eg facial and vocal cues) offers new opportunities for affective computing by incorporating diverse information. Fused representations from text and images have shown improvements over unimodal models for the prediction of sentiment and emotion~\cite{chen2017visual}. As well as this, inter-modality dynamics are harnessed through the integration of multiple modalities, and this has brought forward advancements with respect to sentiment prediction~\cite{zadeh2017tensor, fortin2019multimodal, stappen2019speech}.

Furthermore, the engagement of multimodal data for sentiment analysis has received an increasing level of attention lately~\cite{soleymani2017survey}. The interest of the research community and industry to develop methods, for areas including multimodal sentiment analysis -- which analyse the interaction between users' emotions and topics in multimedia content -- has grown with the wide dissemination of multimodal user-generated content~\cite{zadeh2016mosi, zadeh2017tensor, zadeh2018multi}. While it is established that multimodal approaches lead to higher quality prediction results as compared to unimodal input data~\cite{baltruvsaitis2018multimodal}, due to a lack of robustness, it is still an on-going challenge to develop and employ these techniques in real-world applications. Despite the recent progress for constructing larger datasets~\cite{kossaifi2019sewa} to explore and develop counter strategies for novel `in-the-wild' paradigms, there are many areas which remain unexplored to this day.

In this work, we present in detail the process for the collection and annotation of the \ac{MuSe-CaR} dataset*\footnote[0]{* The data are available online: https://www.muse-challenge.org}. 
The \ac{MuSe-CaR} dataset is a large, extensively annotated multimodal (video, audio, and text) dataset that has been gathered under real world conditions with the intention of developing appropriate methods and further understanding of multimodal sentiment analysis `in-the-wild'. To the best of the authors' knowledge, it is more than three times larger than any other  continuously annotated dataset aimed at pushing the understanding of multimodal sentiment beyond discrete modelling.
Further to this, \ac{MuSe-CaR} provides never before seen annotations, which explicitly allows for modelling of speaker topic and physical entity in relation to continuous emotions.  A selection of the \ac{MuSe-CaR} dataset was utilised for the First International Multimodal Sentiment Analysis in Real-life Media Challenge (MuSe 2020)~\cite{stappen2020muse}, which was held at the ACM Multimedia 2020 conference.

In previous research, multimodal sentiment analysis and emotion recognition are often applied in the context of product reviews~\cite{wollmer2013youtube,morency2011towards}, and sourced from the open-accessible video platform YouTube~\cite{wollmer2013youtube}. With \ac{MuSe-CaR}, we are influenced by this collection strategy, and have designed the dataset with an abundance of computational tasks in mind. Furthermore, the dominant focus for \ac{MuSe-CaR} is to aid in machine understanding of how positive and negative sentiment as well as emotional arousal is linked to an entity and aspects in a review (and other user-generated content in general). In doing so, \ac{MuSe-CaR} aims to bridge fields within affective computing, which currently utilise a variety of emotionally annotated signals (dimensional and categorical). 

We collected over 40 hours of user-generated video material with more than 350 reviews and 70 host speakers (as well as ~20 overdubbed narrators) from YouTube. The extensive annotations consist of 15 different annotation tiers/ tracks (3 continuous dimensional, 3 partially continuous binary, 5 categorical, and 4 automatically extracted tiers). Among others, \ac{MuSe-CaR} offers conversational topic labelling, the novel continuous dimension of \textit{Trustworthiness}, and full word-aligned transcription. 

 When selecting the data, it was of particular importance to balance between the uncontrollable `in-the-wild' influences and constraining properties to allow for meaningful learning and generalisation using current deep learning methods. Such `in-the-wild' characteristics of \ac{MuSe-CaR} include:  \textit{i)}  \emph{video}: face-angle, shot size, camera motion, reviewer visibility, reviewer face occlusion` ((sun-)glasses), and highly varied backgrounds within a single video; \textit{ii)}  \emph{audio}: ambient noises, narrator and host diarisation, diverse microphone types, and speaker locations; \textit{iii)}  \emph{text}: colloquialisms, and domain-specific terms. However, the contextual interaction with emotions, \eg towards entities and aspects, is content dependent. The number of different entities, topics, and aspects that appear in the videos requires special consideration in order to create balanced records for supervised tasks based on vision (image object detection), audio, and linguistics (aspect detection) and their derived fusion. For this reason, we have limited the dataset to vehicle reviews. We consider that reviews from the Automobile domain bring together many of these aspects that are lacking in other review domains (\eg travel, clothing, electronics). For example, the automobile review videos show half-covered faces, different soundscapes (e.g., engine, wind) simultaneously as the reviewers articulate product descriptions and their opinions.
 Furthermore, most of the reviewers are semi- or professional reviewers (\eg YouTube channels, influencers). This has several practical advantages: on the one hand, it increases the video quality significantly, on the other hand, it makes the videos more consistent (a broad but similar range of topics around the vehicle is covered, \eg vehicle safety). 
Finally, approval to use multiple videos is greatly simplified, as a channel with multiple videos (and independent speakers) only needs to be contacted once and not per video\footnote{We contacted the creators for consent, see \Cref{sec:selection} for details.}. 

Based upon this dataset, our contributions in this work are:

\begin{itemize}
    \item First, we give in-depth information regarding the \ac{MuSe-CaR} selection, collection, and annotation process, which was not addressed in the MuSe 2020 challenge baseline paper\cite{stappen2020muse}. This includes the presentation of additional annotation tiers that have not been utilised and introduced, yet (\cf \Cref{sec:tiers}). We expect that this will assist future participants and other researchers to conduct and interpret studies on \ac{MuSe-CaR} more easily.
    \item Second, we revisit the tasks introduced in MuSe 2020 and demonstrate that the limits in terms of performance in, at least one of these tasks, has not been reached, yet. When doing so, by proposing a simple, yet efficient model which utilises state-of-the-art components we beat the baseline of MuSe-Trust by around $50$ \%. Furthermore, we describe extensive experiments run to identify the key settings in modelling the novel task of Trustworthiness.
\end{itemize}
Given the novelty of the described dataset, and the extensive analysis that has already been made by the authors, there are numerous ideas and further directions which could be taken by researchers in the community. With this in mind, we outline an array of research directions which make use of various combinations of the data annotation tiers.

\vspace{-1em}
\section{Related Resources}\label{sec:related}

In the following, we highlight important databases for \ac{MuSe-CaR} to build upon which focus on computational sentiment and emotion analysis from audio-visual recordings. For an overview of databases which utilise only one or other combinations of modalities in this area, the reader is referred to recent survey studies (\eg for text~\cite{karas2020deep}, for vision~\cite{li2020deep}, and for  audio\cite{swain2018databases}). An overview of findings based on our specific areas of interest are summarised in \Cref{tab:databaseoverview}.

\begin{table*}[t!]
\centering
\small
\caption{Selection of multimodal sentiment analysis and affective computing datasets focusing on at least one of three types of prediction targets: (Sentiment) classes, (Primitive) Emotions, Object-of-Interest.
\textbf{Modal}ities available;
Language: MULTI 1 = CN, DE, EN, GR, HU, SE \& MULTI 2 = DE, ES, FR, PT;
\textbf{An}notation \textbf{Du}ration (hh:mm)\protect\footnotemark; 
\textbf{\#} Number of minimum \textbf{Anno}tations per target. Subjectivity includes
Sentiment: \textbf{\#} Number of \textbf{sent}iment classes (* derived sentiment intensity from Valence); \textbf{\#} number of (basic) \textbf{Emo}tions; \textbf{Cont}ionous,
Primitive: \textbf{Dim}ensions: \textbf{V}alence, \textbf{A}rousal, \textbf{T}rustworthiness, \textbf{L}ikability, \textbf{I}ntensity, \textbf{P}ower, \textbf{E}xectiation, \textbf{D}ominance; \textbf{\#} number of \textbf{Inc}rement \textbf{St}eps,  \textbf{T}race annotations;
\textbf{O}bject-\textbf{o}f-\textbf{I}nterest: classes of topics or entities.
}
\label{tab:databaseoverview}
\resizebox{.6\linewidth}{!}{%
    \begin{tabular}{lccrccccccc}
        \toprule
        \multirow{2}{*}{\textbf{Name}} & \multirow{2}{*}{\textbf{Modal}} & \multirow{2}{*}{\textbf{Language}} & \multirow{2}{*}{\textbf{AnDu}} & \multirow{2}{*}{\textbf{\# Anno}} & \multicolumn{2}{c}{\textbf{Sentiment}} & \multicolumn{3}{c}{\textbf{Primitive}} & \multirow{2}{*}{\textbf{OoI}}     \\
        & & & &  & \# Sent & \# Emo &  Class  & \# IncSt  & Cont &  \\       \hline
        MuSe-CAR  & V,A,L & EN & 40:12 & 5 & 5* & \xmark & V,A,T & \xmark & \cmark & \cmark \\ \hline
        \multicolumn{11}{c}{\textbf{Affective Computing}} \\ \hline
        Aff-Wild~\cite{kollias2019deep}& V,A & EN & \footnotemark30:00 & 6 & \xmark & 6 & V,A & \xmark & \cmark & \xmark \\
        SEWA~\cite{kossaifi2019sewa} & V,A & MULTI 1 & 4:39 & 5 & \xmark & \xmark & V,A,L & \xmark & \cmark & \xmark \\
        HUMAINE~\cite{douglas2007humaine} \ & V,A & EN & 4:11 & 6 & \xmark & \xmark & V,A,I & 7 & \cmark & \xmark \\
        RECOLA~\cite{ringeval2013introducing} & V,A & FR & 3:50 & 6 & \xmark & \xmark & V,A & 9 & \cmark & \xmark \\
        AFEW-VA~\cite{kossaifi2017afew} & V,A & EN & 2:28 & \xmark & \xmark & \xmark & V,A & 21 & \xmark & \xmark \\ \hline
        VAM~\cite{grimm2008vera} & V,A & EN & 12:00 & 6-8 & \xmark & 5 & V,A & 5 & \xmark & \xmark \\
        IEMOCAP~\cite{busso2008iemocap}& V,A,L & EN & 11:28 & 5 & \xmark & 9 & V,A,D & 5 & \xmark & \xmark \\ 
        SEMAINE~\cite{mckeown2011semaine} & V,A & EN & 6:30 & 6 & \xmark & 7 & V,A,I,P,E & \xmark & \cmark & \xmark \\ 
        Belfast~\cite{sneddon2011belfast}& V,A & EN & 3:57 & 6 & \xmark & \xmark & V,A & 3 & \xmark & \xmark \\ \hline
        \multicolumn{11}{c}{\textbf{Multimodal Sentiment}} \\ \hline
        UR-FUNNY~\cite{hasan2019ur}& V,A,L & EN & 90:23 & 2 & \xmark & 1 & \xmark & \xmark & \xmark & \xmark \\ 
        MOSEAS~\cite{zadeh2020moseas} & V,A,L & MULTI 2 & 68:49 & 3 & 7 & 6 & \xmark & va & \xmark & \xmark \\
        MOSEI~\cite{zadeh2018multimodal} & V,A,L & EN & 65:53 & 3 & 7 & 6 & \xmark & \xmark & \xmark & \xmark \\ 
        ICT-MMMO~\cite{wollmer2013youtube} & V,A,L & EN & 13:58 & 2 & 5 & \xmark & \xmark & \xmark & \xmark & \xmark \\ 
        Ext. POM~\cite{marrese2020multi} & V,A,L & EN & 15:40 & 1 & 5 & \xmark & \xmark & \xmark & \xmark & \xmark \\ 
        CH-SIMS~\cite{yu2020ch} & V,A,L & CN & 2:20 & 5 & 5 & \xmark & \xmark & \xmark & \xmark & \xmark \\ 
        AMMER~\cite{cevher2019towards} & V,A,L & DE & 1:18 & 1 & \xmark & 5 & V,A & 11 & \xmark & \xmark \\ 
        Youtubean~\cite{marrese-taylor-etal-2017-mining} & V,A,L & EN & 1:11 & 2 & 3 & \xmark & \xmark & \xmark & \xmark & \xmark \\
        MOUD~\cite{perez2013utterance}& V,A,L & ES & 0:59 & 2 & 3 & \xmark & \xmark & \xmark & \xmark & \xmark \\
        YouTube~\cite{morency2011towards} & V,A,L & EN & 0:29 & 3 & 3 & \xmark & \xmark & \xmark & \xmark & \xmark \\ \hline
        \bottomrule
    \end{tabular}
     \vspace{-1em}
}
\end{table*}
\footnotetext{Note: In the case of multimedia sentiment analysis databases, we have usually indicated the total size. However, the exclusion of non-spoken parts typically reduces the size significantly, and authors tend not to specify the size of the adjusted audio-visual database. For example, 33 hours of audio-visual data were collected for the SEWA database, but only 14\,\% of the corpus are annotated with emotions. For Aff-Wild, only the total size of the dataset but the extend of annotation is not explicitly stated.}

\vspace{-1em}
\subsection{Multimodal sentiment analysis datasets}
It is generally accepted~\cite{liu2012survey, mantyla2018evolution} that (multimodal) sentiment analysis consists of a holder and the object (subject, entity) that the emotion is evoked from. 
Furthermore, the survey of~\cite{soleymani2017survey} divides the field into three major groups: multimodal sentiment analysis \textit{i)} in (monologue) video reviews \eg from video platforms~\cite{wollmer2013youtube,rosas2013multimodal,zadeh2020moseas}, \textit{ii)}  in human-machine and human-human interactions~\cite{kossaifi2019sewa}, \textit{iii)} analysing of general multimedia content (\eg images, gifs) from social media~\cite{borth2013large}, and stresses the need for additional datasets to extend the field. 
Recently, \textbf{UR-FUNNY}~\cite{hasan2019ur} collected a large number of Ted-Talk videos on more than 400 topics. On a small part of this data, binary humour is predicted. \textbf{MOSEAS}~\cite{zadeh2020moseas} is a multilingual collection of 40\,000 audio-visual sentences consisting of sentiment, emotion, and attribute labels. \textbf{MOSEI}~\cite{zadeh2018multimodal} contains videos of 250 topics for sentiment analysis (7 classes) and emotion recognition (6 classes). Videos with no transcriptions and punctuation are provided by the creator, additionally they exclude video content where the camera was not fixed in place.~\cite{garcia2019multimodal} extended \textbf{POM}~\cite{park2014computational}, an audio-visual film review dataset, with annotations at the level of opinion-forming segments and components. It consists of 600 videos with an average length of 94 seconds in which a person looks straight into the camera and talks about six film aspects. \textbf{ICT-MMMO}~\cite{wollmer2013youtube} consists of user-generated review videos to predict the sentiment. \textbf{CH-SIMS}~\cite{yu2020ch} acquired 60 Mandarin raw videos from movies, and television series and shows with segments of up to 10 seconds. It is limited to parts where both, the face and the voice appears at the same time. \textbf{Youtubean}\cite{marrese-taylor-etal-2017-mining} collects seven popular product reviews of one cell phone model with aspect and sentiment annotations. \textbf{MOUD}~\cite{perez2013utterance} includes \textit{YouTubers} clearly visible (no occlusions) and oriented frontally to the camera expressing their opinions in 30 second segments without any background music. The \textbf{YouTube} corpus~\cite{morency2011towards} provides sentiment labels for a wide range of YouTube product reviews. \textbf{AMMER}~\cite{cevher2019towards} focuses on emotional interactions in a simulated car journey in German.
Discrete emotions used in these databases are not fully eligible to represent the sentiment, making more complex representations~\cite{soleymani2017survey} \eg through polarity and intensity scores necessary.

\vspace{-1em}
\subsection{Affective computing datasets}
One such way to represent the sentiment more comprehensively is through the primitive dimensions of emotions, \eg the circumplex model of emotions~\cite{russell1980_Circumplex}. In this context, Valence often serves as an umbrella term of sentiment and is used interchangeably~\cite{thelwall2010sentiment, mohammad2016sentiment,preoctiuc2016modelling}.
\textbf{Aff-Wild}~\cite{kollias2019deep} consists of 30 hours of audio-video material on emotion recognition (valence, arousal) focusing on facial images in a variety of head poses, lighting conditions and occlusions. It was later expanded ~\cite{kolliasexpression} by an additional 260 YouTube videos with a total length of about 13 hours, annotated under similiar conditions and at frame level adding action units and 7 basic expressions.
\textbf{SEWA}~\cite{kossaifi2019sewa} provides a large audio-visual data with continuous arousal, valence, and likeability traces during human-human interactions recorded online via static webcams. However, only around 4 hours are continuously annotated. \textbf{HUMAINE}~\cite{douglas2007humaine} provides continuous intensity annotations in addition to Arousal and Valence. The \textbf{RECOLA} dataset~\cite{ringeval2013introducing} contains subjects interacting in a tightly controlled laboratory environment. The audio, visual, and electro-dermal activity were annotated with Arousal and Valence traces.
\textbf{AFEW-VA} includes 7 facial expressions from films, annotated by 3 raters, in addition to Arousal and Valence with intensities from -10 to 10 hard incremental steps.
\textbf{VAM} consists of clips from a German talk-show. Besides Valence, Arousal, and dominance, which are annotated on a discretised 5-point scale, there are also six basic emotions which are sparsely labelled. \textbf{IEMOCAP}~\cite{busso2008iemocap} consists of audio, video, and transcriptions of ten actors in dyadic sessions in a controlled setting. \textbf{SEMAINE}~\cite{mckeown2011semaine} is a richly annotated database of 21 human-agent interaction sessions. \textbf{Belfast}~\cite{sneddon2011belfast} is a collection of actively stimulated participants evoke moderate emotional response due to specific tasks. The recordings come with continuous values of Valence and Arousal annotations. 

\vspace{-1.5em}
\subsection{Summary}\label{sec:litsum}
From this overview of recent literature, on the one hand, we find that multimodal sentiment analysis databases currently try to select content from a wide range of topics, which increases the (linguistic) generalisation potential of developed models. However, multimedia databases which utilise topics or aspects as a prediction target~\cite{marrese-taylor-etal-2017-mining,garcia2019multimodal} are rare, limiting the (supervised) understanding of the context almost exclusively to linguistic analysis. 
To date, no such methods offer complete solutions and rely on the spoken word. In situations where spoken language is not present and towards multimodal understanding, we want to examine the relationships between emotion and object/ physical entity. As the literature suggests a) these topic definitions, \eg reviews\cite{zadeh2018multimodal} are too high-level to justify an in-depth \textit{understanding} of the opinion-topic structure and their multimodal interactions, a point which is necessary for real-life applications, and b) with the drastic improvements of general language models (\cite{floridi2020gpt}) generalisation improves naturally over time~\cite{wang2019glue}.

On the other hand, with a wide range of (health and wellbeing) situations in mind, affective computing focuses on the elementary sensing of affects using often primitive and generalistic (continuous) emotional dimensions, making it very broadly applicable~\cite{luneski2010affective}. These enable short- and long-range understanding and aggregation of affects, emotions, and sentiments. However, understanding what these are aimed at (\eg subject, entity) is not necessarily seen as part of the field. Moreover, the shift towards perception in a noisy environment (in-the-wild), especially with regard to visual characteristics, has only recently been made a focus~\cite{kossaifi2019sewa, kossaifi2017afew}. We argue that a) with the improvements of models through deep learning and its efficient utilisation on large-scale data, this focus might change, and b) the properties of continues traces might be helpful when dynamically breaking down a large sequence of audio-visual emotional events into shorter segments,  \eg sentences, aspects or noun-adjective pairs in the future.  

The \ac{MuSe-CaR} dataset described in this work was designed to overcome some of these basic limitations when utilising user-generated, real-life media on a large-scale and attempts to bring the best of two worlds together. We consider the following aspects in the design of \ac{MuSe-CaR} listed below.
\begin{itemize}
\item The recording of the database should be as uncontrolled to a high degree in terms of recording settings, emotional and linguistic content as possible. However, there should be a certain overlap in the topics and aspects addressed.
\item Instead of isolated segments and sentences, the database should contain long sequences across multiple aspects of a topic.
\item The emotion-object interaction of the speaker should be in various environments (\eg outside and inside of a car). 
\item The depicted speaker is not actively acting emotions, but rather, emotions are naturally elicited depending on the topic, aspect and situation.
\item The audio-visual material provides verbal and nonverbal information. However, occasionally, one modality provides only limited information (voice but no face, \eg  the camera focuses on an object; face but no voice, \eg acceleration; audio-visual but no transcription, \eg speech-to-text failed due to complex audio scenarios).
\item The emotional and topic annotations should be assigned based on human subjective evaluations.
\end{itemize}

\section{The \texorpdfstring{\ac{MuSe-CaR}}~~database} \label{sec:selection}
In the following, we describe how we identified relevant material based on the previously defined criteria. Furthermore, we outline the communication process with the creators to reach consent for the use of their content. Then, we provide information regarding the dataset composition, as well as the data annotation process and the utilised annotation tools. 

\vspace{-1em}
\subsection{Data acquisition}
The selection of videos from YouTube was carried out in a semi-automatic process. We developed a basic crawler which receives a number of hand selected keywords (\eg `review' and a car brand) and provides metadata of pre-selected videos. In our view, the legal situation for crawling videos from the web is inconclusive in many countries\footnote{Uploading a video to YouTube automatically issues that video under the YouTube own license. Regarding this licence, the use of the data in the EU is only possible by YouTube directly or with the consent of the creator. In similar works~\cite{zadeh2016mosi,zadeh2020moseas,zadeh2018multimodal}, the database producers refer to the fair use principle for academic use. These exceptions of intellectual property rights, however, do not seem to be applicable in the European legal sphere. Furthermore, YouTube's standard terms and conditions as well as of those from the API have to be considered. A fraction of videos are also available under the Creative Commons (CC-BY, full use if the creator credits are mentioned) licence model.}. Therefore, we contacted the creators of the videos with high user engagement (views, likes, etc.) which most likely indicates relevance and high-quality content. Requesting actively the creators' consent to use the data for academic purposes in an opt-in approach, gives researchers worldwide legal certainty when using the database. We sent up to three (follow-up) emails to the creators over a period of three months, reaching an agreement in around 50\,\% of requests. In the emails, we explained the intention of the dataset for non-commercial use in challenges and for research and provided an example of an End User License Agreement (EULA).  

\vspace{-1em}
\subsection{Data inspection and selection}
If the consent was given, three individuals carried out a deeper initial inspection by viewing around 10\,\% of each video (one assessment per video). In total, 366 videos were inspected. Based on the criteria defined in \Cref{sec:litsum} and the content of each video, the inspectors filled out a survey, asking to estimate important data properties such as the level of in-the-wild characteristics. These also included the emotionality and quality of the video for our purpose (\cf meet the derived criteria in \Cref{sec:litsum}) on a scale of 1 to 5 (where 1 is substandard, and 5 is optimal).  For example, a substandard video would rather have shots of a car but is not a review or the majority of the filming took place in a studio. Regarding emotionality, less than 5\,\% are rated below 3 and around 80\,\% as 4 or 5. Furthermore, regarding video quality properties, they seem suitable for our purpose with more than 85\,\% holding scores equal to or higher than 4. For instance, purely neutral videos with no emotional interaction, \eg without a human, synthetic speech narration listing criteria of the car.
Furthermore, additional estimation questions reveal more detailed information about the videos (\eg face inclusion): shot range, or camera selfie-angles, the scene and noise settings (including the quality of sound), whether there are background music and other speakers present, and if speakers have a dialect or accent. These additional estimation questions, are specifically general, and the annotators prior to observing did not receive detailed instruction on their definition, other than being given the ‘template’ video which was given as a ‘gold standard’ In regards to nativeness and dialect, although we do not have a ground truth for this, and we notes that particular non-native speakers may have a less obvious accent, we consider consider that deeper analysis that a ‘first-impression’ may be out of scope. 

\begin{figure}
\centering
\includegraphics[width=0.8\linewidth]{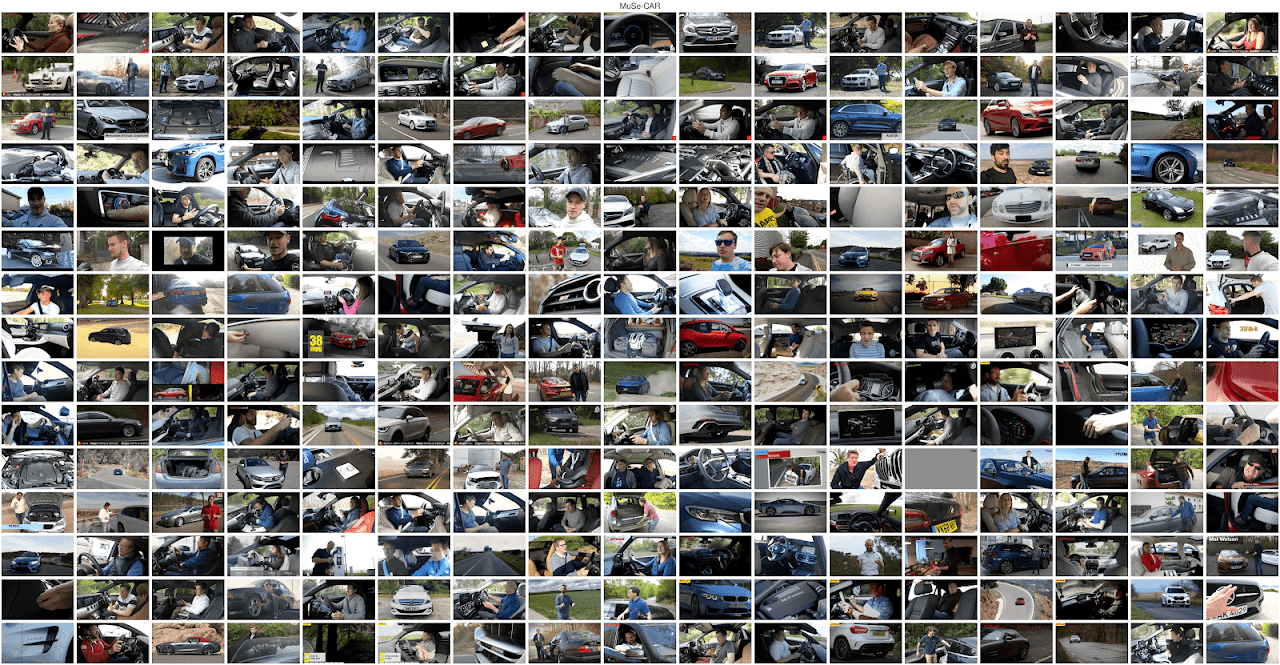}
\caption{Thumbnails showing reviewers in various constellations to the camera and interacting with the object.}
\vspace{-1em}
\label{fig:thumbnails}
\end{figure}

\begin{figure}
 \centering
 \includegraphics[width=0.49\linewidth]{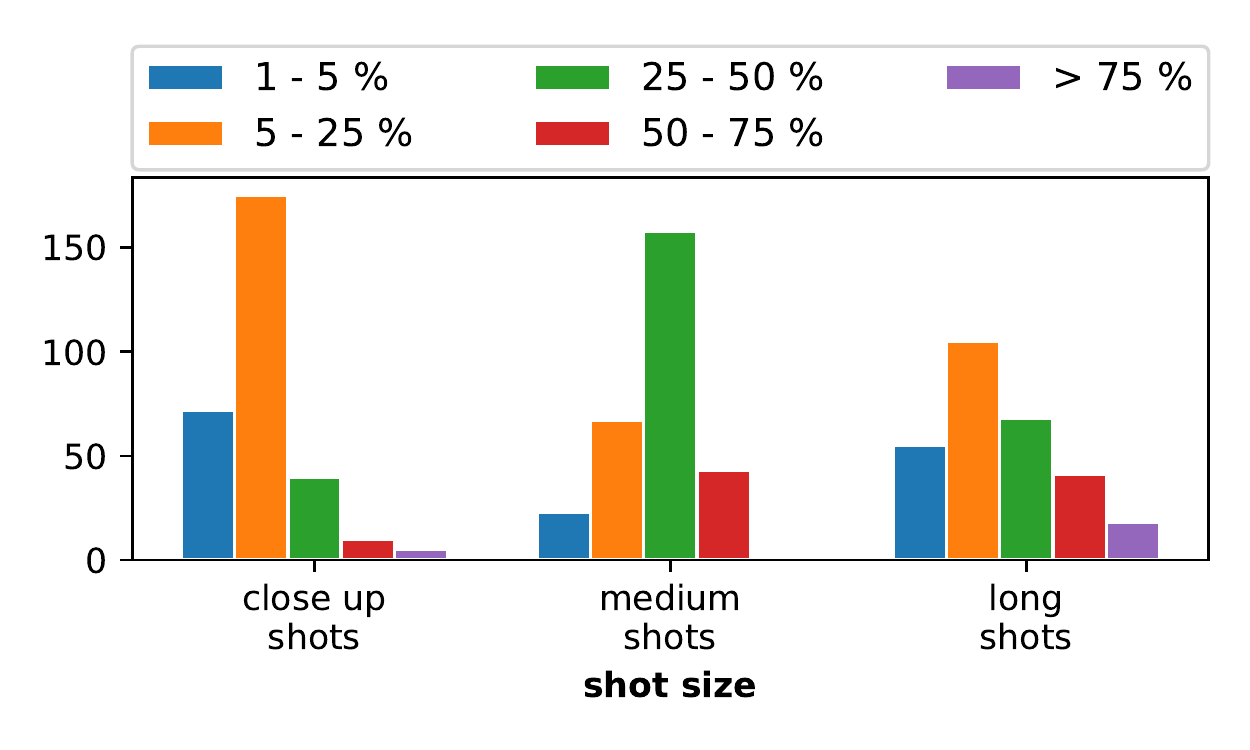} 
 \includegraphics[width=0.49\linewidth]{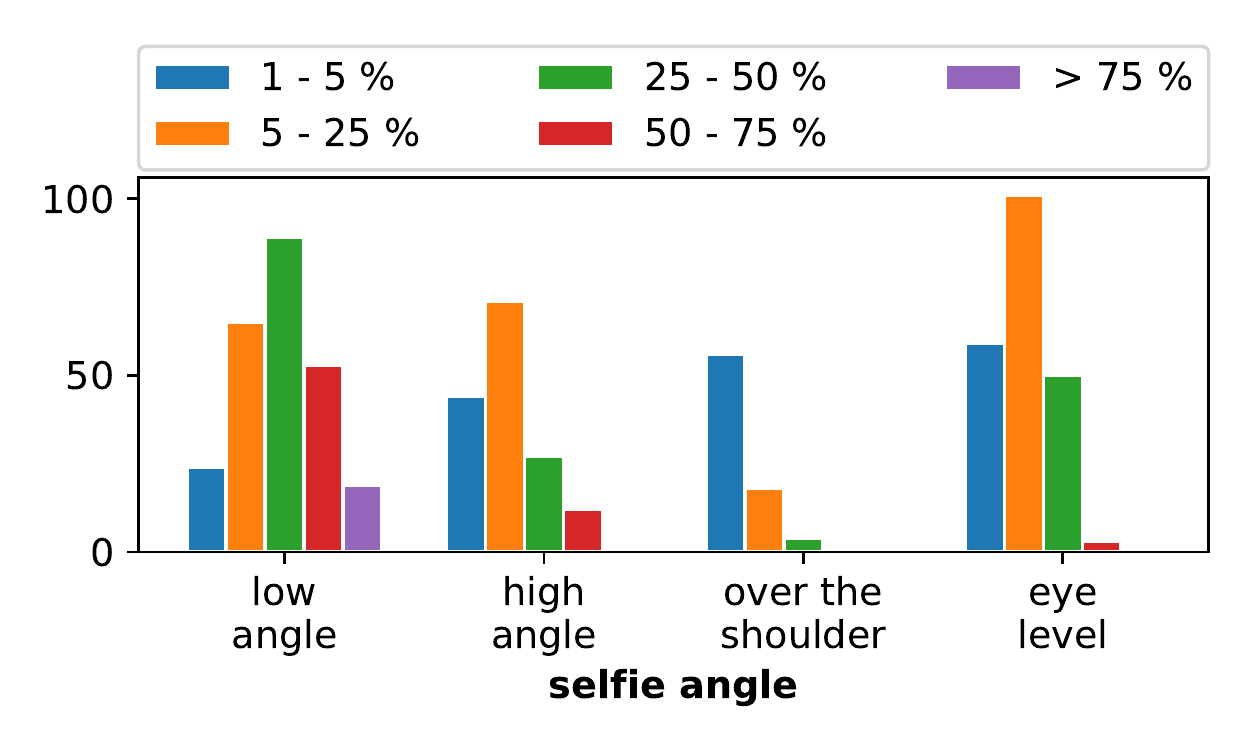} \\
 \includegraphics[width=0.49\linewidth]{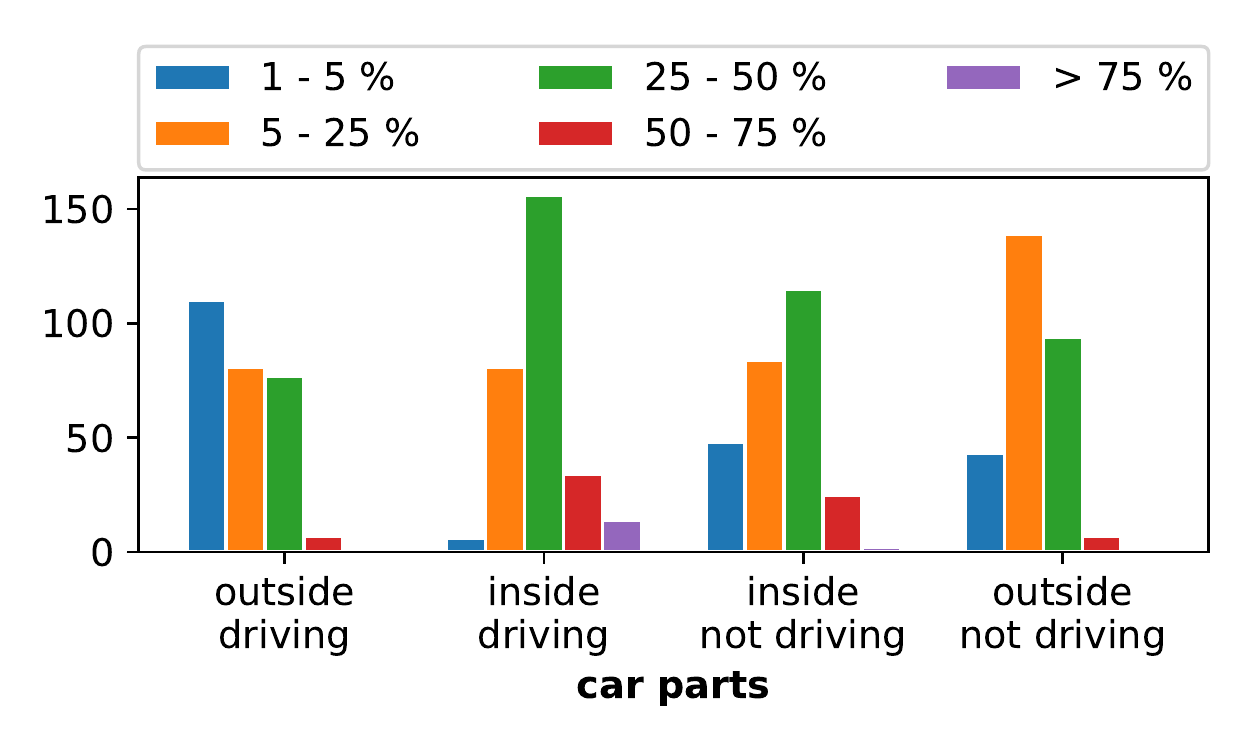}
 \includegraphics[width=0.49\linewidth]{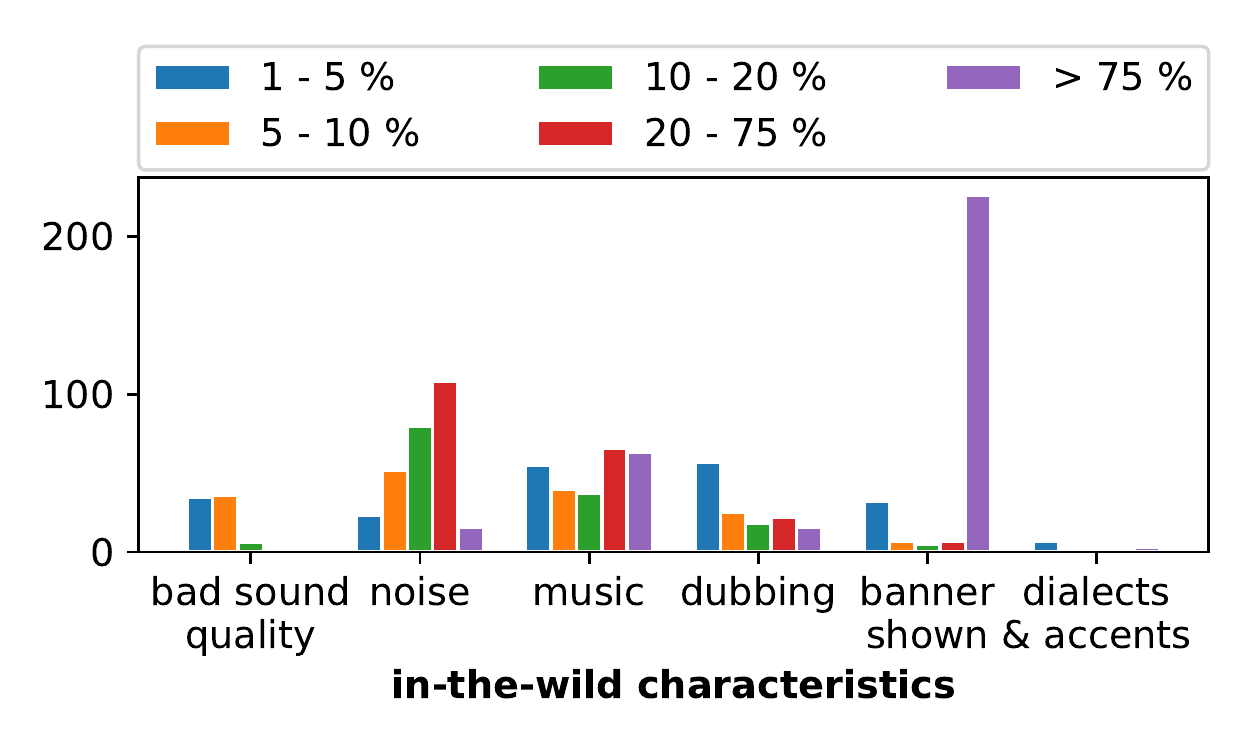} 
 \caption{Absolute counts of the estimation of various in-the-wild characteristics, namely camera shot size, camera selfie-angle, scene setting, and additional noise influences of the collection. The percentage estimates the duration of collected video in a certain category, \eg 25-50\,\% in ``close up shots'' means that the collector estimates that 25-50\,\% of the duration of the collected video contains close up shots. 
 }
 \label{fig:camera}
 \vspace{-1em}
\end{figure}

A detailed analysis can be seen in \Cref{fig:camera}. Interesting observations are that there are fewer close up than medium sized shots in most videos (only accounts for 1-25\,\% of the videos, \cf \Cref{fig:thumbnails}). For selfie shots, the camera is mainly held at a lower angle. Videos are often filmed inside the car, either while driving or not driving. Fewer coverage accounts for outside-the-car shots (for both driving and not driving). In general, there are noise sounds and background music present, however, the sound quality is only perceived to be bad for 1-10\,\% of the video duration. The great majority of videos has a banner present (\eg copyright sign in a corner) for over 75\,\% of the video duration. 

Since the subjects were not actively selected, but professional, semi-professional (`influencers'), and casual reviewers, we can only estimate the characteristics of our cohort. We assume a broad age range from the mid-20s until the late-50s while most speakers are English natives from the United Kingdom or the United States of America; a small minority are non-native, yet fluent English speakers. 
Around a third of reviewers wear glasses. There are barely any videos with speakers having a dialect or accent. 

Videos which received less than 4 points of overall quality for training, less than 3 on emotionality or lack of the properties outlined in the previous section were excluded. After inspection, 303 videos remained for annotation, which corresponds to 40.2 hours of video with an average duration of 8 minutes (90\,\% are shorter than 14 minutes).




\vspace{-1em}
\subsection{Post processing: voice-activity, transcription \label{sec:transcripts}}
It is well established that linguistic information of the spoken language can facilitate the learning of emotions and is a cornerstone for understanding context. To support future research into the interplay of the audio, visual, and text modalities, we automatically transcribed the data. In recent years, speech-to-text achieved almost human-level quality in popular languages, such as English. Furthermore, using the text modality for emotion recognition in a real-life scenario would need to work independently, without human intervention, and considering the size of the dataset along with these prerequisites, we decided to use automatic transcription services on our corpus.

The transcriptions from the videos using Google Cloud speech API\footnote{https://cloud.google.com/speech-to-text} and Amazon Transcribe\footnote{https://aws.amazon.com/transcribe/} were both of sufficient quality. The first contains also non-verbal cues and audio elements, such as laughter, music, theme, etc. 

\begin{figure*}[ht]
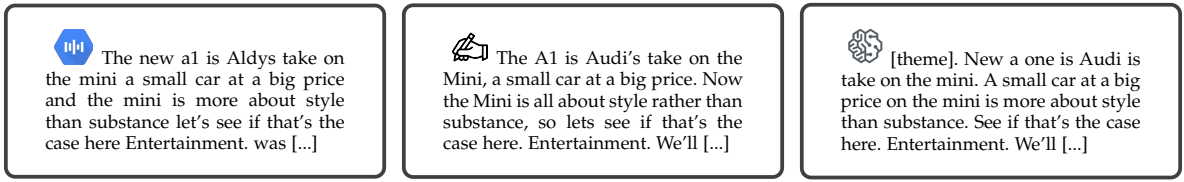

\centering
 \begin{minipage}{.28\linewidth}
  \google{\scriptsize The new a1 is Aldys take on the mini a small car at a big price and the mini is more about style than substance let's see if that's the case here Entertainment. was [...]} 
 \end{minipage}%
 \hspace{0.1cm}
 \begin{minipage}{.28\linewidth}
   \hand{\scriptsize The A1 is Audi's take on the Mini, a small car at a big price. Now the Mini is all about style rather than substance, so lets see if that's the case here. Entertainment. We'll [...]}
 \end{minipage}%
  \hspace{0.1cm}
 \begin{minipage}{.28\linewidth}
  \aws{\scriptsize [theme]. New a one is Audi is take on the mini. A small car at a big price on the mini is more about style than substance. See if that's the case here. Entertainment. We'll [...]} 
 \end{minipage}%
 \vspace{-5pt}
   \caption{The video (ID: 265) with the worst speech-to-text result from our hand-transcribed (middle) selection with a word error rate of 37.85\,\% from Google-Transcribe (left) and 39.44\,\% from AWS (right).}
  \label{fig:trans} 
   \vspace{-10pt}
\end{figure*}

Ten videos were randomly selected for human transcription to estimate the word-error rate (WER). The total number of transcribed words is 10 576. The WER of Google Cloud speech API in this sample corresponds to 25.04\,\%, while that of Amazon Transcribe is 28.39\,\%. In \Cref{fig:trans} we have shown an example of a section from the video with the highest error rate from our sample. This analysis indicates that even if the values seem objectively high, typical errors are often minor such as ``A1 is Audi's'' (hand) vs ``a1 is Aldys'' (google) vs ``a one is Audi'' (AWS). We provide both, but only use the latter for our experiments since we felt that the word timestamps and car-specific vocabulary slightly exceeds the first one. One reason for this improvement may be the option to create a customised dictionary to improve the transcription quality regarding domain-specific, automotive typical terms. 

The transcriptions of the spoken language includes punctuation (\eg period, question mark, exclamation mark) and every transcribed word comes with a beginning and an end timestamp as well as duration. These metadata help to align the text with the annotations (different sampling rate) and other modalities as well as enabling studies on more than 28\,295 sentences exceed all English Multimodal Sentiment Analysis databases (\cf \Cref{sec:related}) and have almost 5k more sentences than the next biggest (MOSEI).

Although we have not rigorously evaluated them, our preliminary screening suggests that the punctuation and boundaries are accurate and even better than commonly available voice activity detectors we also have tried.


\vspace{-1em}
\subsection{Data annotation}

\subsubsection{Annotation roles and organisation}
The size of the dataset and fine-grained annotations required a highly efficient annotation process. In order to ensure high quality, ethical and meaningful annotations for the \ac{MuSe-CaR} dataset, we considered that keeping a human in the loop was a vital aspect~\cite{baird2020considerations}. We therefore defined three functional roles:

\begin{enumerate}
\item \textbf{Annotator}: It is the responsibility of the annotator to label the data based on the subsequent instruction.
\item \textbf{Auditor}: It is the responsibility of the auditor to review the performance of the information labelled, and ensure it is in line with the annotation protocol. Only after the annotations have been manually and automatically checked, verified and endorsed by the auditors is the annotation deemed usable.
\item \textbf{Administrator}: Manage all parties, and assign duties during the entire annotation process.
\end{enumerate}

During annotation, an interactive process between the annotator, auditor, and administrator was applied:

\begin{enumerate}
\item \textbf{Assignment of tasks:} The annotator is assigned one or more packages by the administrator. Similar to previous work~\cite{busso2008iemocap}, a session package corresponds to ca.\ 40 minutes of video material. The annotators were instructed to have suitable rests between videos and sessions, so that the expected (and paid) working time was one hour. In one session, all videos had to be annotated with the same annotation type (\eg Valence).  The videos are distributed in two rounds. In the first round, three annotators annotate a package and after all of them have finished, round two takes place, where the remaining 2 annotations per package are carried out. This division of allocation is necessary to determine the quality in round one (1 vs 2 annotators) and that no imbalance occurs by improving the quality of late assigned packages. In other words, with equal distribution, early assigned packages have a worse quality than those assigned at the end, see also quality tracking).
\item \textbf{Annotation}: The annotator annotates the videos piece by piece and package by package and  sends packages to the auditor after completion. Once all packages have been processed and evaluated by the auditor, new packages can be assigned to the annotator.
\item \textbf{Progress tracking:} The administrator regularly tracks the progress of the annotators and auditors. This keeps track of which packages are still to be annotated or audited -- in the worst case, reevaluating the suitability of the work load. 
\item \textbf{Quality tracking:} By calculating the similarity (Concordance Correlation Coefficient (CCC)) of the annotations in a “one vs all” strategy of the batch-wise assigned videos, high disagreements can be identified.  These are then manually investigated regarding qualitative aspects by the Auditor (see above) according to the annotation protocol. The Auditor provides detailed feedback to the annotators and with continuous auditing throughout, the quality of the annotator can be tracked and the quality is therefore continuously improved.
\end{enumerate}

All annotators hold at least a Bachelor's degree, while the minimum requirement for the auditors and administrators was a Master's degree in a technical field and at least 2 years of work experience related to the research field.

\vspace{-.5em}
\subsubsection{Annotation tools}
\begin{figure}
    \centering
    \includegraphics[width=0.8\linewidth]{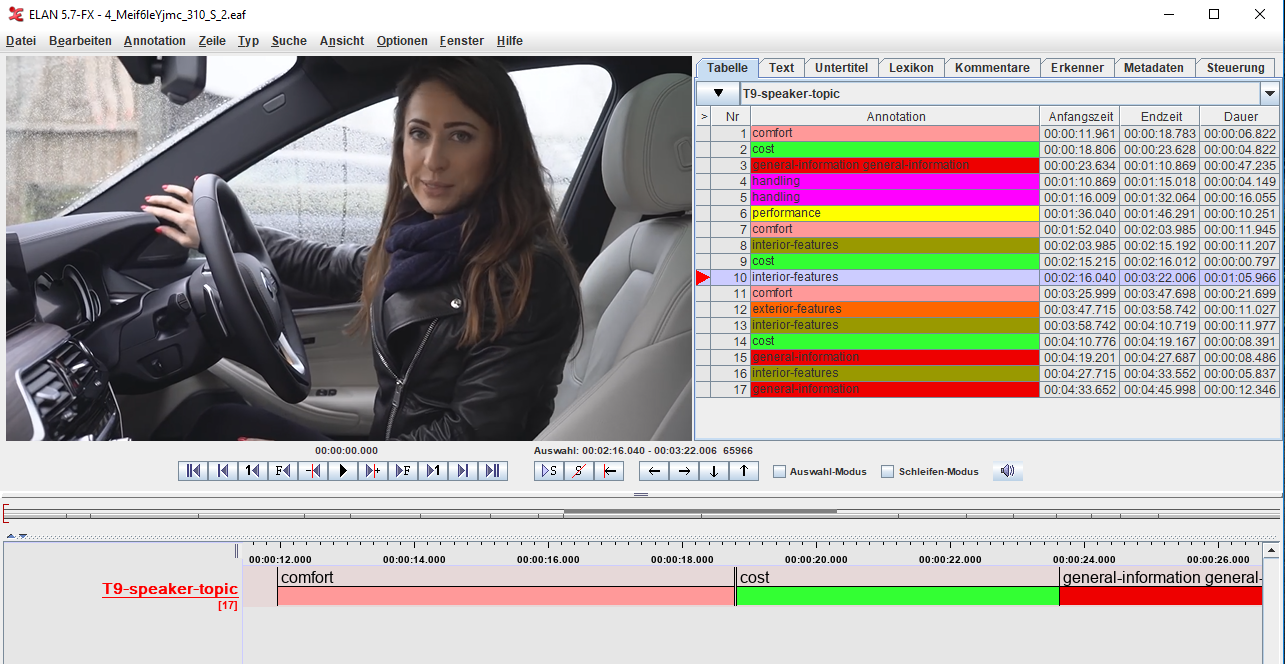}
    \vspace{-0.4cm}
    \caption{Interface of the annotation software ELAN when labelling speaker topics (\cf \Cref{sec:speakertopics}).}
    \label{fig:elan}
    \vspace{-1em}
\end{figure}

For categorical annotation (\eg speaker topic), we used the annotation software ELAN 4.9.4\cite{wittenburg2006elan} -- chosen for its multimodal interface which allows for a waveform and video display (see \Cref{fig:elan}), as well as other useful functionalities including the ability to jump to areas of interest~\cite{wittenburg2006elan}.

It was shown that some emotions are transmitted more strongly via visual signals (\eg sadness) while others more via audio (\eg anger)~\cite{de1997facial}. In addition, context information transported by both modalities plays a crucial role in emotion perception~\cite{barrett2011context}. For an audio-visual annotation of the continuous emotions, we choose the software DARMA~\cite{girard2018darma}. DARMA enables to record annotation signals from a \textit{Logitech Extreme 3D Pro Joystick}. The joystick allows for the transfer of perceived emotions more intuitively~\cite{kossaifi2019sewa}. The continuous annotation was made from the very start to the end of a video and was sampled at 0.25\,Hz with an axis magnitude of range -1\,000 and 1\,000.  In other words, while an annotator was watching a video, the joystick was simultaneously moved in real-time depending on the perceived emotion, i.e., downwards when the valence changes from positive to negative and the joystick signal are recorded by DARMA.

\vspace{-1em}
\subsection{Annotation tiers}\label{sec:tiers}

\begin{figure}
    \centering
    \includegraphics[width=\linewidth]{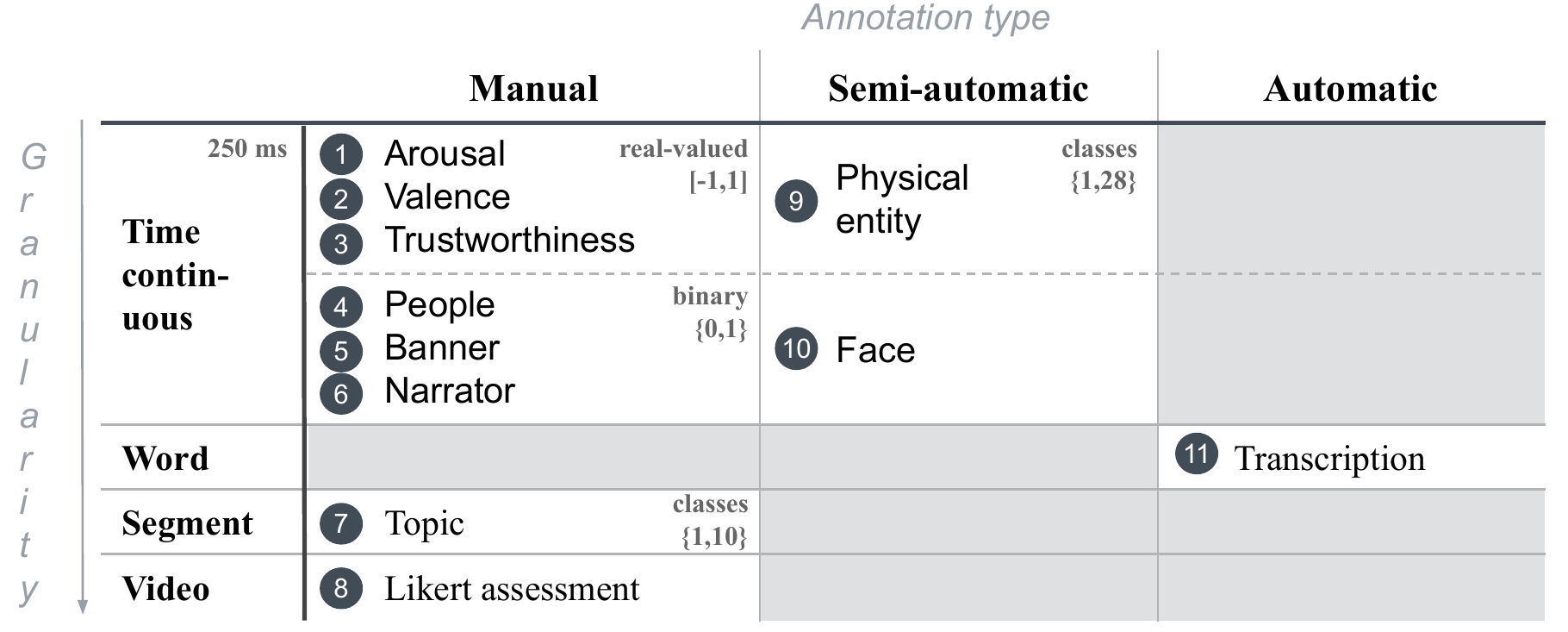}
    \vspace{-0.4cm}
    \caption{Overview of the type and granularity (sampling rate) of all our annotations: 1-3: continuous real-valued annotations (\cf \Cref{sec:a_v} and \Cref{sec:t}); 4-5: continuous binary annotations (\cf \Cref{sec:people}, \Cref{sec:banners}, and \Cref{sec:narrhost}); 9: physical car entity classes (\cf \Cref{sec:gocard}); 10: faces (\cf \Cref{sec:face}); 11: transcriptions (\cf \Cref{sec:transcripts}); 7: speaker topic segments (\cf \Cref{sec:speakertopics}); 8: likert assessment (\cf \Cref{sec:likert}).}
    \vspace{-0.4cm}
\end{figure}

The \ac{MuSe-CaR} contains annotations for continuously-valued (Valence, Arousal, and Trustworthiness), binary-valued (host/narrator turns, banner, and person appearance), and categorical (topics, entities) ratings. Overall, we have annotated 11 tiers for each video. The dimensional annotation reflects the continuous emotional state of the individuals speaking.

In parallel to the continuously-evaluated signals, simple continuous binary (activated/deactivated) annotations are recorded simultaneously by pressing and holding the trigger adjacent to the index finger of the joystick: 
\textit{i)} Trustworthiness + the turns between the host and the narrator, 
\textit{ii)} Valence + the appearance of banners, and 
\textit{iii)} Arousal + the appearance of more than one person.

In addition to the annotations, we provide more than 10 pre-computed features,  \eg facial landmarks, acoustic low-level descriptors (LLDs), hand gestures, head gestures, facial action units, etc., which in other works~\cite{stappen2019speech, kossaifi2019sewa} were also decelerated as (semi-)automatic generated annotations for prediction. A detailed description of these can be found in~\cite{stappen2020muse}.

\begin{figure}
 \centering
 \includegraphics[width=0.45\linewidth]{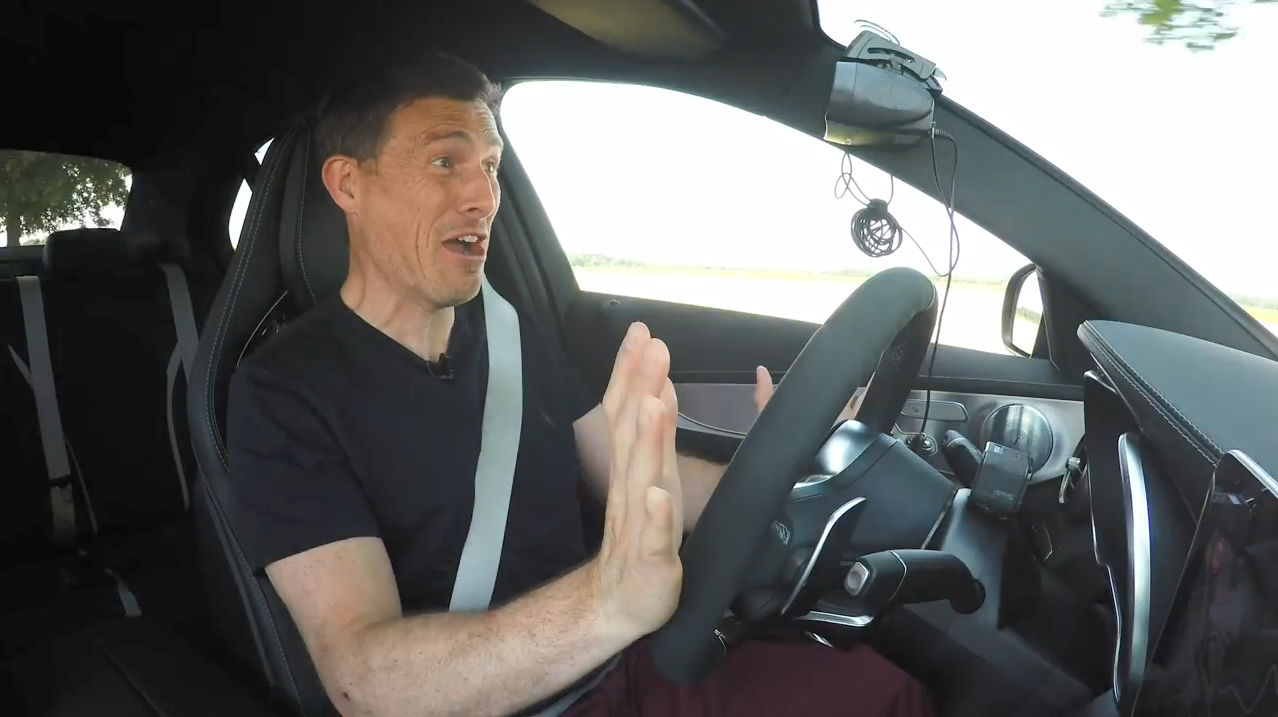} 
 \includegraphics[width=0.45\linewidth]{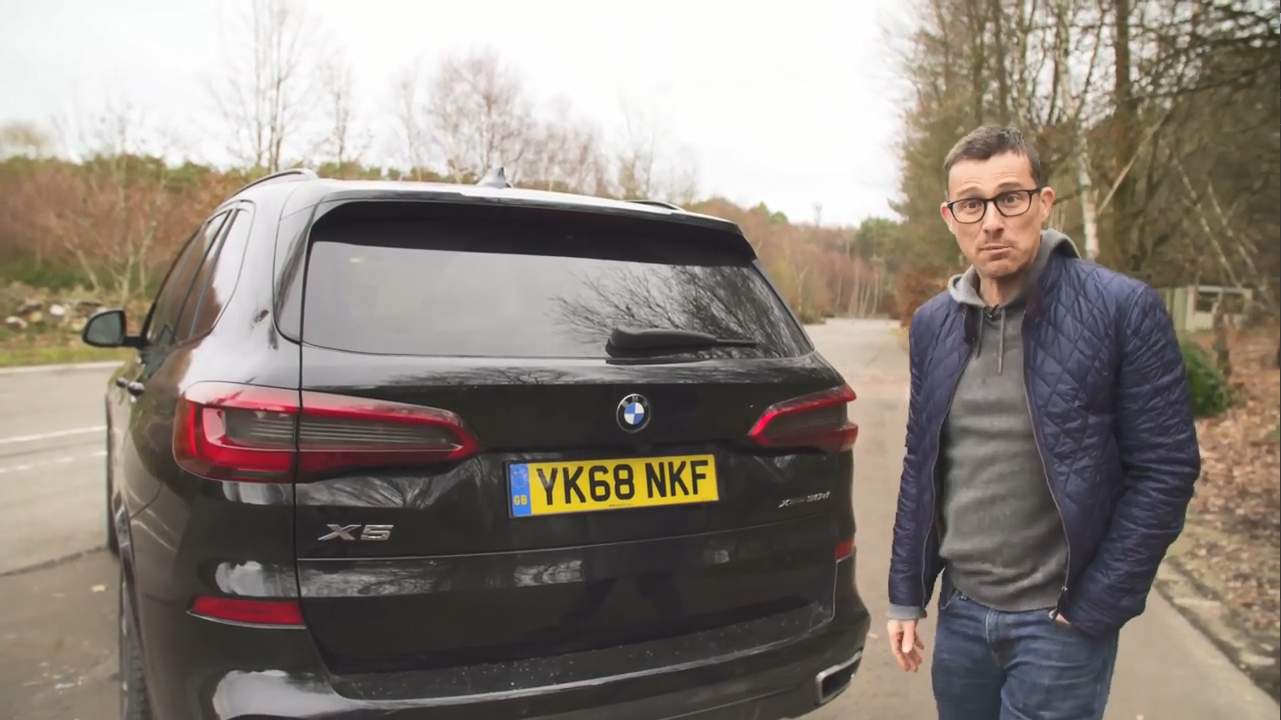} 
 \caption{Snapshot of emotional scenes from the \ac{MuSe-CaR} dataset depicting high Arousal (left) by non-verbally expressing the excitement about the acceleration of the car and verbally positively valenced appreciation of the cars' exterior.}
  \vspace{-10pt}
 \label{fig:highvalue_examples}
\end{figure}

\vspace{-.5em}
\subsubsection{Arousal and Valence\label{sec:a_v}}\index{T1 Arousal and T2 Valence}
The videos are annotated using a continuous dimensional model of emotion~\cite{russell1980_Circumplex}, considering both Arousal and Valence (\cf \Cref{fig:highvalue_examples}). We discuss these together for ease, however, the literature shows that it is important to annotate them separately~\cite{kossaifi2019sewa}. In the dataset, an example of high Arousal is \emph{Stressed} or \emph{Elated (happiness)}, however \emph{Stressed} would be negative Valence, and \emph{Elated (happiness)} would be positive Valence.

The annotators were trained in-person. First, the axes were introduced by showing them an explanatory video 
to impart a general understanding of aspects of emotion. After further explanation and examples, they could experience the handling of the software and the reaction of the joystick in a hands-on session. The test annotations were followed by a group discussion, where individual annotations were compared to an up-hand recorded annotation of an experienced annotator as well as between the group members. The final packages were done individually in a quite environment with headsets. 

\subsubsection{Trustworthiness\label{sec:t}}
User-generated information has proven useful when creating large (emotional) datasets with real-world content~\cite{zadeh2018multimodal}. It is known~\cite{schwemmer2018social} that to build up reputation of a -- in real life unknown -- creator has a strong influence to the user engagement and, thus, also most likely on other perceived emotions. However, quantifying Trustworthiness is hard~\cite{moturu2011quantifying}, and there is yet no dataset that offers the possibility to link it to Arousal and Valence as well as train cross-domain detectors. 

Since this is a completely novel dimension, we explain our definition more deeply. Generally there is no single, prevailing definition of Trustworthiness due to the lack of a conceptual agreement~\cite{horsburgh1961trust, moturu2011quantifying, cox2016trustworthiness}. Analogous to our understanding, in~\cite{colquitt2007trust},  Trustworthiness is defined as the ability, benevolence, and integrity of a trustee.

In the context of a stranger (our moderator) from social media content, Trustworthiness presupposes that this person can objectively assess the (facts of the) matter on the one hand and on the other hand communicates this assessment unbiased, therefore honestly. Building on this definition, we asked the annotators to evaluate the Trustworthiness throughout the video, \ie when the host is discussing a particular aspect, how honest and knowledgeable does the annotator feel their review is? In other words, this could also include the annotator perceiving a commercial gain rather than a truthful review of the product. 

Several video examples and cases were given to the annotators. Although, we reiterate that trustworthiness can be a subjective tier for annotation, and therefore additionally consider this from the annotator's perspective: Do you believe the information that is given to you? Do you have the feeling their argumentation is based on facts and experience or is it that they are rather trying to sell something.

\subsubsection{Dimensional annotation fusion}\index{Dimensional Annotation Fusion}

Several methods exist to fuse a set of subjective emotion annotations to establish a consensus from individual annotations~\cite{panagakis2015robust, schuller2013intelligent}. Since there can be no fully objective signal of subjective information such as emotions, this label is referred to as the gold-standard.

\begin{figure}
    \centering
    \includegraphics[width=.6\linewidth]{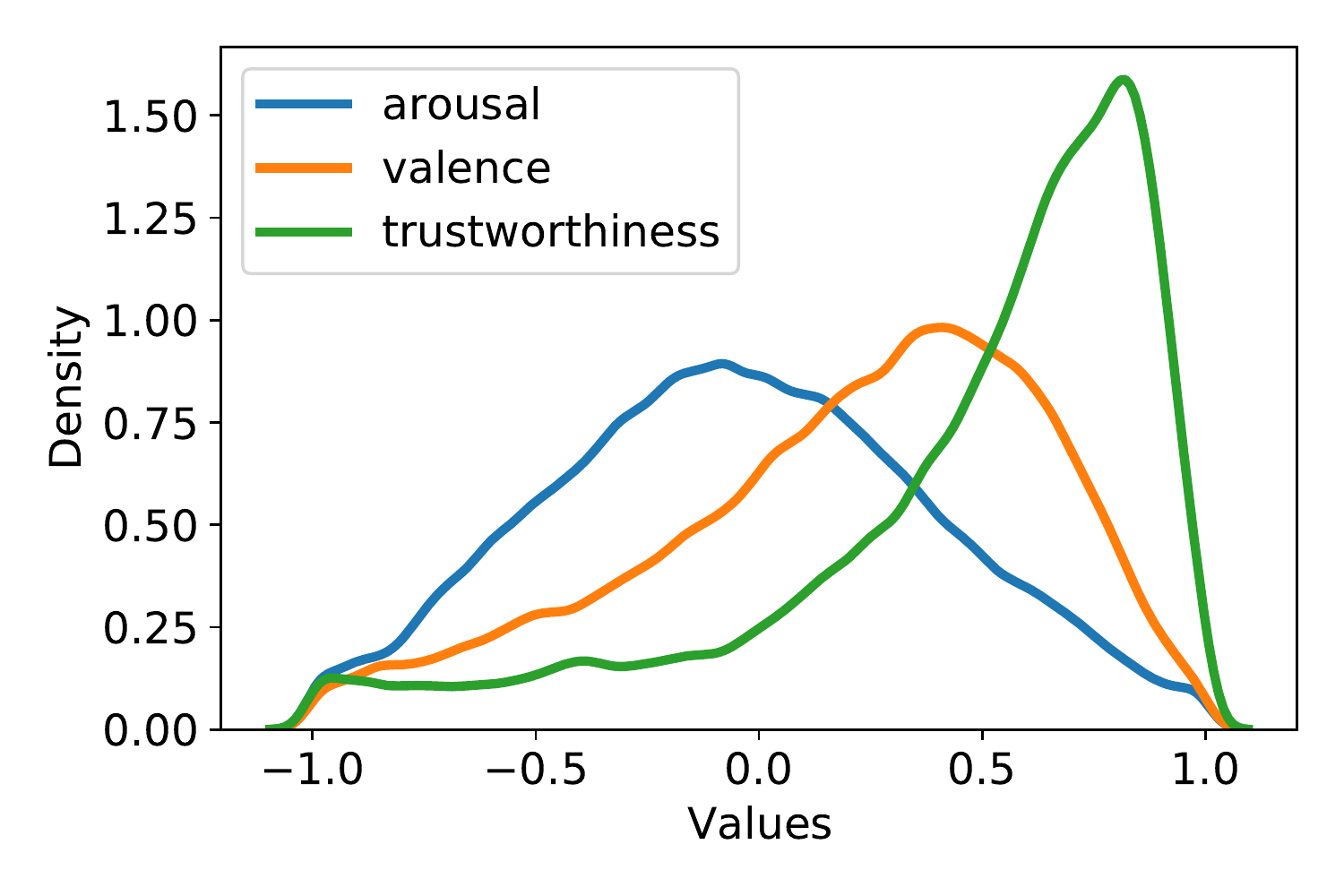}
    \vspace{-0.4cm}
    \caption{Density estimation of the continuously-annotated dimensions Arousal, Valence, and Trustworthiness tiers after EWE fusion of individual annotator tracks. While Arousal is almost perfectly Gaussian-shaped, Valence is skewed toward the positive end of the spectrum, and Trustworthiness is even more peaked.}
    \label{fig:freq}
    \vspace{-0.4cm}
\end{figure}

\Cref{fig:freq} depicts the frequency distribution of the created gold-standards utilising the \ac{EWE} approach~\cite{schuller2013intelligent}. Essentially, \ac{EWE} considers the reliability of each annotation by calculating the cross-correlation of the annotation with the mean annotation (over all annotators). It can also be seen as the weighted mean of the annotators' agreement~\cite{grimm2005evaluation, hantke2016introducing}. \ac{EWE} is described further in~\cite{schuller2013intelligent}, and has been applied to multiple similar continuous emotion databases~\cite{ringeval2013introducing, ringeval2017avec,kossaifi2019sewa}.

For \ac{MuSe-CaR}, every video is annotated by at least five independent annotators employed by the EIHW Chair from a group of eleven (six female and five male), all fluent in English. The age of the annotators ranges from 21 to 30 years. In this context, the mean \textit{concordance correlation coefficient (CCC)} is used to measure the inter-rater agreement across all annotations for each dimension is given as: Arousal $.265$, Valence $.350$, and Trustworthiness $.316$. The levels of agreement are moderate as to be expected~\cite{douglas2005multimodal} and consistent with those of other emotional datasets\cite{kossaifi2019sewa, devillers2005challenges} showing that the stronger `in-the-wild' characteristics seem not to have a strong influence on the perception of emotions. We would like to note that the trustworthiness dimension in \cref{fig:freq} is strongly left-skewed. Additional research is required to understand whether the underlying cause of this phenomenon, \eg our data source (YouTube), domain, or selection approach. 

\begin{figure}
    \centering
    \includegraphics[width=1\linewidth]{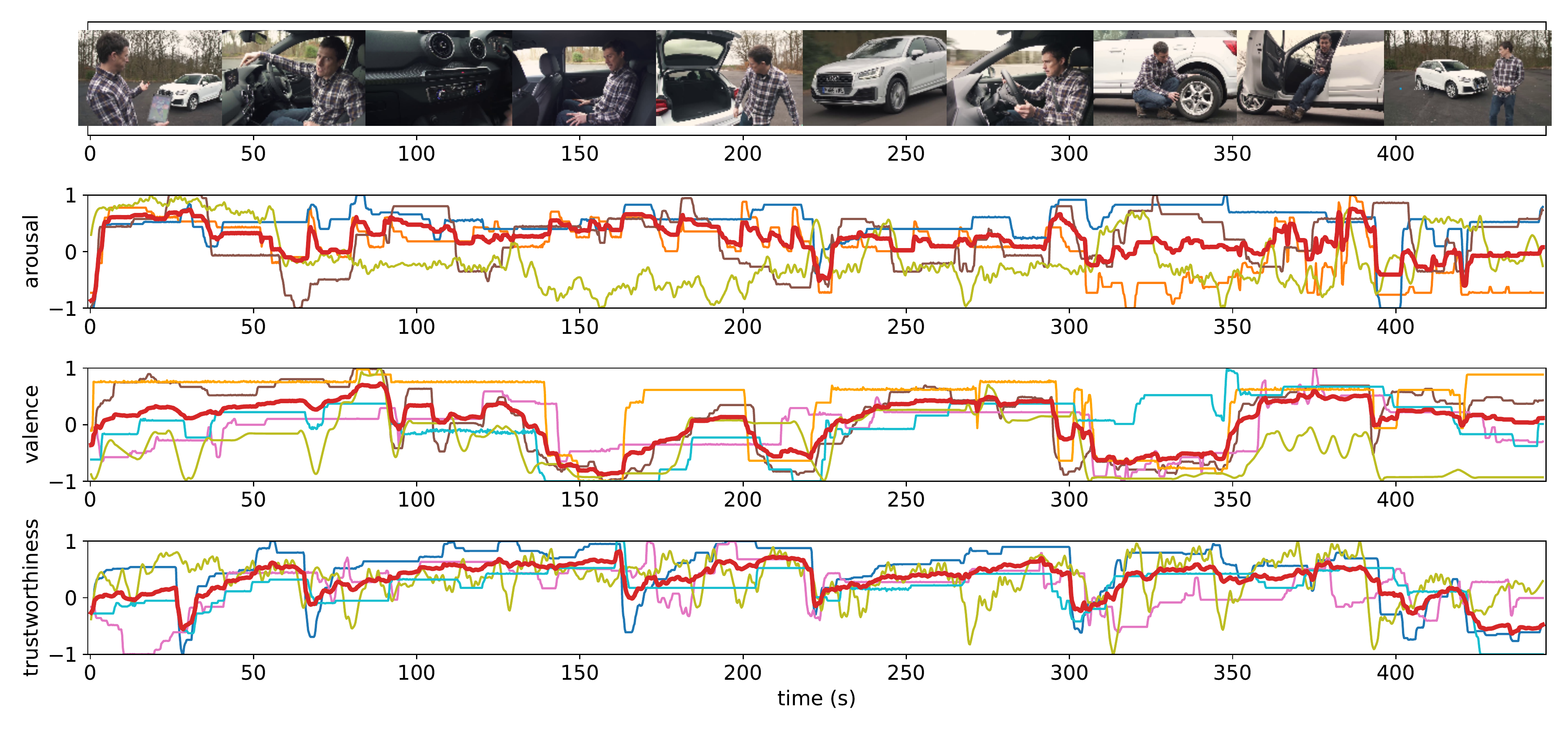}%
    \vspace{-0.4cm}
    \caption{Annotation example (video id: 236) for the three continuous, real-valued dimensions (valence, arousal, trustworthiness) with a sample rate of 250ms showing the single annotations and the EWE fusion (bold, red).}
    \label{fig:timeline}
    \vspace{-0.2cm}
\end{figure}
\vspace{-1em}

\subsubsection{Host \& narrator turns} \label{sec:narrhost}

Within the videos, there are two human subjects which are of main interest to the annotators; these 
have been defined as follows: 
\begin{itemize}
    \item The \textbf{host} is the reviewer/presenter of the car and its features, talking often directly to the audience and expressing own opinions. 
    \item The \textbf{narrator} presents in some videos, may occur in the opening and closing sequences, may provide additional information, and is not visible to the audience.
\end{itemize}
During the annotation of Trustworthiness, the turns between the \textbf{host} and the \textbf{narrator} were annotated using the binary trigger. When the host is not speaking, and the narrator speaks, the binary trigger is held.

\subsubsection{Banners\label{sec:banners}}
Social media networks, such as YouTube, are a large, self-extending and diverse data pool. In these videos, however, superimposed graphics occur such as text and channel logos -- some for copyright reasons, others to either inform or entertain the viewer. For our tasks, such banners might not be useful, since they might obscure entities or the host. Therefore, visible banners that show up are annotated in parallel to Valence. In a later stage we want to detect, and measure the influence on visual features or exclude/ replace them~\cite{slucki2018extracting, tian2015text, yang2016scenetextreg}.

Banners appear with a variety of properties: \textit{i)} appearance: static, or dynamic, \textit{ii)} positions: bottom-right, bottom-left, upper-right, upper-left, entire footer, centre, or changing, \textit{iii)} timing: on-screen billing, opening statement, in-between, or closing credits, \textit{iv)} duration: highlight (very short), short (few seconds), or consistent (copyright), and \textit{v)} transparency: none, partly, or mostly transparent.

\subsubsection{Multiple people\label{sec:people}}

The focus of this database is on videos with only one visible speaker/ person. In a few cases, more than one person may appear, either for specific comments or for demonstration purposes, \eg how many people fit in the back seat, or the host is being shown a novel feature. These situations are binary annotated in parallel to Arousal.

\subsubsection{Speaker topic}\index{T9 Speaker Topic}\label{sec:speakertopics}
Speaker Topics rely on generalisable topics vocalised by the host and related to the object of interest. To obtain the topic labels, we started by selecting a broad range of the most mentioned nouns from automatically extracted transcriptions. This list of elements was then reduced manually by watching a random selection of 20 videos from the dataset, gradually gaining knowledge of more frequent topics, based on speaker interests and expressions.
These higher-level topics combine many different aspects under one term. The videos are labelled by speaker topic segments while often one topic segment compromises of several sentences (\cf \Cref{fig:elan}). A detailed overview of sub-topics and aspects covered by each topic can be found in \Cref{tab:overview_speakertopics}. It also shows the distribution of the topics across all 28k sentences, while 20\,\% have more than one topic annotated.

\begin{table}[ht!]
\caption{Annotated speaker topics with examples of sub-topics and aspects. The \% gives the share of the number of sentences in each topic, while 20\,\% of sentences have multiple topics.}
\vspace{-0.4cm}
\label{tab:overview_speakertopics}
\resizebox{\columnwidth}{!}{%
\begin{tabular}{ll|l|ll|l}
\toprule
\multicolumn{3}{c|}{\textbf{Feature exterior (7\,\%)}} & \multicolumn{3}{c}{\textbf{Feature interior (6\,\%)}} \\ \hline
light & \multicolumn{2}{l|}{headlight, foglight, taillight} 
& audio system & \multicolumn{2}{l}{radio, speaker} \\
door exterior & \multicolumn{2}{l|}{locks, handle} 
& seat & \multicolumn{2}{l}{belt, split folding backs} \\\hline

\multicolumn{3}{c|}{\textbf{Handling, Driving Experience (13\%)}} & \multicolumn{3}{c}{\textbf{User Experience (7\,\%)}} \\ \hline
driving actions & \multicolumn{2}{l|}{braking, steering, gear shifting} 
& infotainment & \multicolumn{2}{l}{\begin{tabular}[c]{@{}l@{}}screen, bluetooth,
realtime traffic  
 \end{tabular}}  \\
dynamics & \multicolumn{2}{l|}{centroid, chassis, suspension}
& interaction & \multicolumn{2}{l}{interface, iDrive system, gestures} \\ 
\hline
\multicolumn{3}{c|}{\textbf{Performance (13\,\%)}} & \multicolumn{3}{c}{\textbf{Quality \& Aesthetic (7\,\%)}} \\ \hline
powertrain & \multicolumn{2}{l|}{electric, hybrid, combustion} 
& design & \multicolumn{2}{l}{interior, exterior,  style (sporty, etc.)} \\
engine & \multicolumn{2}{l|}{\begin{tabular}[c]{@{}l@{}}horsepower, RPM, acceleration  \end{tabular}} 
& quality & \multicolumn{2}{l}{material quality, clearance} \\
\hline
\multicolumn{3}{c|}{\textbf{Safety (2\,\%)}} & \multicolumn{3}{c}{\textbf{Comfort (6\%)}} \\ \hline
tests & \multicolumn{2}{l|}{Euro NCAP, NHTSA, rating} 
& surface & \multicolumn{2}{l}{leather, touch} \\
assistance sys.\ & \multicolumn{2}{l|}{\begin{tabular}[c]{@{}l@{}}anti-lock brakes, traction control\end{tabular}} &
space & \multicolumn{2}{l}{leg room, head room, luggage} \\
\hline
\multicolumn{3}{c|}{
\textbf{General information (16\,\%)}} & \multicolumn{3}{c}{\textbf{Costs (3\,\%)}} \\ \hline
introduction & \multicolumn{2}{l|}{\begin{tabular}[c]{@{}l@{}}series, weight, sales, warranty\end{tabular}} 
& one-off & \multicolumn{2}{l}{retail price, base price, feature price} \\
comparison & \multicolumn{2}{l|}{models, brands, competitors} 
& after sale & \multicolumn{2}{l}{insurance, maintenance, resale} \\ 
\bottomrule

\end{tabular}%
}
\end{table}

\subsubsection{Physical entities}\index{Entity}\label{sec:gocard}
In contrast to the voice-based speaker topics, the physical entities rely purely on visual input. The domain-specific objects of human-object interaction, and car parts, were annotated with bounding boxes for all recordings with a step size of 4 frames per second. 28 types of exterior and interior car parts, such as, door, steering wheel, and infotainment were labelled. 

Manual bounding box annotation of all frames, each with a high number of classes, are highly labour intensive. Based on previous experiments~\cite{su2012crowdsourcing, konyushkova2018learning}, we can assume that around two boxes per minute can be labelled. Since there are 576\,000 frames to annotate with up to 15 boxes for each frame, it would be impractical to label all of them manually (requiring between 4\,800 and 72\,000 hours for a single annotator).
\begin{figure}
    \centering
    \includegraphics[width=\linewidth]{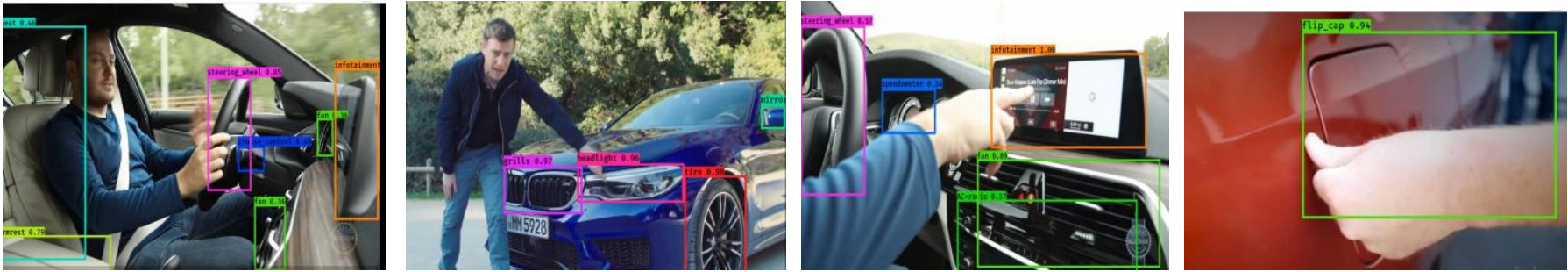}
    \vspace{-0.4cm}
    \caption{Examples of interaction between the host and the car parts from GO-CarD~\cite{stappen2020go}.}
    \label{fig:gocard}
    \vspace{-0.4cm}
\end{figure}

We chose a semi-automatically process. First, a localiser (Darknet-53 network) is pre-trained on 15\,003 vehicle images from other datasets. This underlying data does not include any human interaction. Therefore, we extracted and labelled another 1\,000 frames compromising of more than 8\,000 boxes from \musec and fine-tuned the algorithm. A detailed description can be found in a separate paper~\cite{stappen2020go}. The network achieves a mean average precision of 67.6\,\% with scores up to 94.0\,\% for very distinctive parts. A visual inspection was performed on the annotations and those deemed unsatisfactory were removed or corrected. \Cref{fig:gocard} shows examples of these parts during human-object interactions. 

\subsubsection{Face extraction\label{sec:face}}\index{Face Extraction}
For the same reasons as in \Cref{sec:gocard}, we applied a semi-automatic process to the extraction of faces, the start (start of visible face) and end (end of visible face) point of occurrence as well as the relative position within a frame. 
We labelled faces in a small selection of videos from each channel by hand, to measure quantitatively the success of our automatic extraction and localisation. \mtcnn~\cite{zhang2016joint} provides a robust framework for this task, previously already proved to be accurate for face annotation in an emotional context~\cite{kossaifi2019sewa}. It has a cascaded structure of three stages and is trained on the datasets WIDER FACE~\cite{Triantafyllos} and CelebA~\cite{liu2015faceattributes}. 

We classified the detected bounding boxes into true and false positives resulting in an accuracy of $90$\,\%, and an F1 score of $86$\,\% on our selection. Furthermore, we conducted a visual inspection of the bounding boxes. Given the high level of visual in-the-wild characteristics, for instance, partly visible faces, different sizes, side-shots, sunglasses, etc., we consider both, quantitative and qualitative results as strong and sufficient for further feature extraction.

As described in~\cite{stappen2020muse} in detail, we use frameworks such as VGGFace~\cite{parkhi2015deep}, and OpenFace~\cite{baltruvsaitis2016openface} to extract face-related features including intensity and presence \acp{FAU}, facial landamrks, head pose, and gaze position.

\subsubsection{Additional annotation-related metadata\label{sec:likert}}
Each dimensional and categorical annotation is followed by a very brief survey of the annotator's perception of the content viewed and a self-assessment of their own annotation. On a 10 point Likert scale (0 not at all, 10 very much) annotators are asked four questions: 
 \textit{i)} How appealing did you find the video? 
 \textit{ii)} How emotional did you find the host? 
 \textit{ii)} How trustworthy did you find the content? 
 \textit{vi)} How confident are you about the accuracy of your annotation?

\begin{figure}
 \centering
 \includegraphics[width=0.49\linewidth]{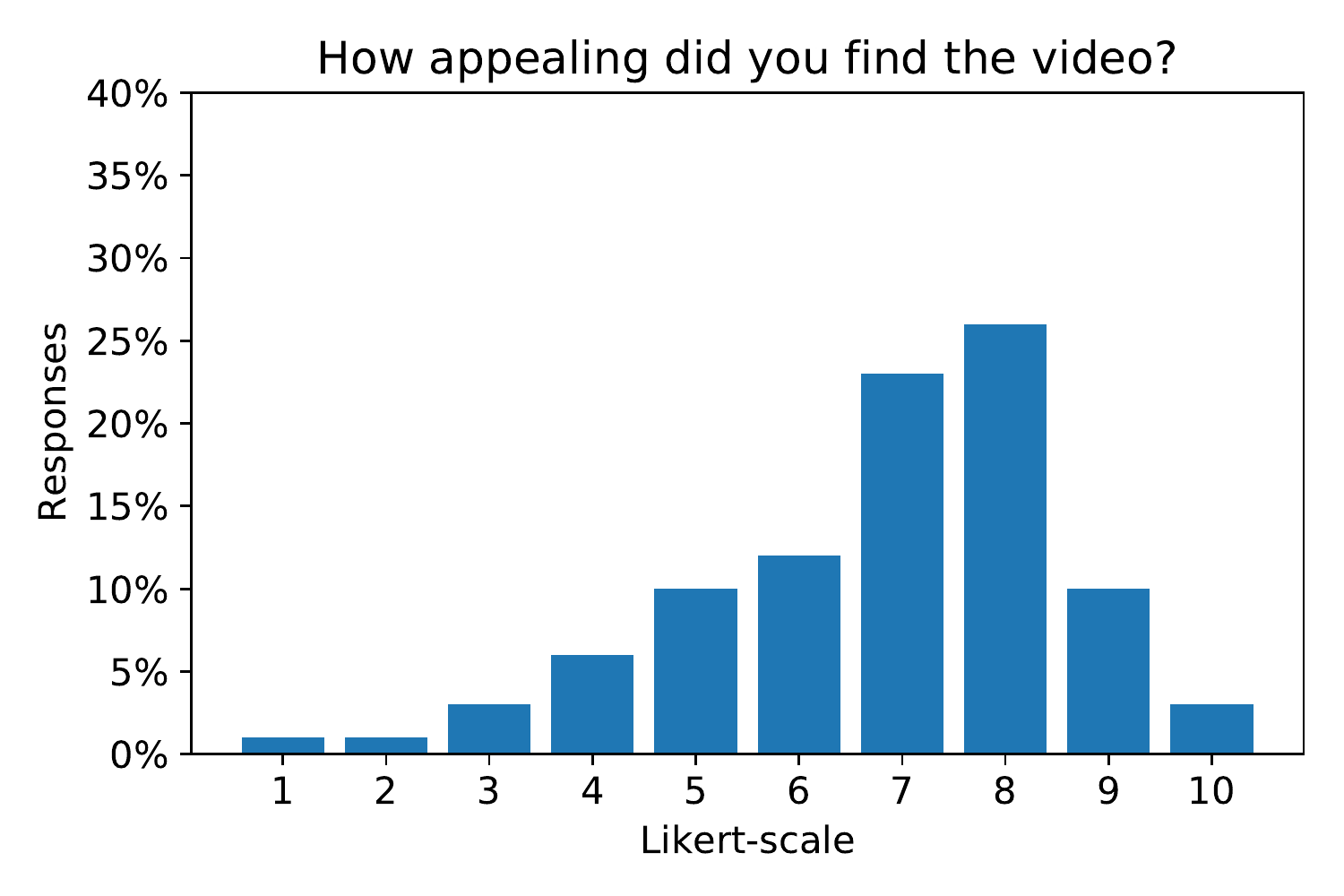} 
 \includegraphics[width=0.49\linewidth]{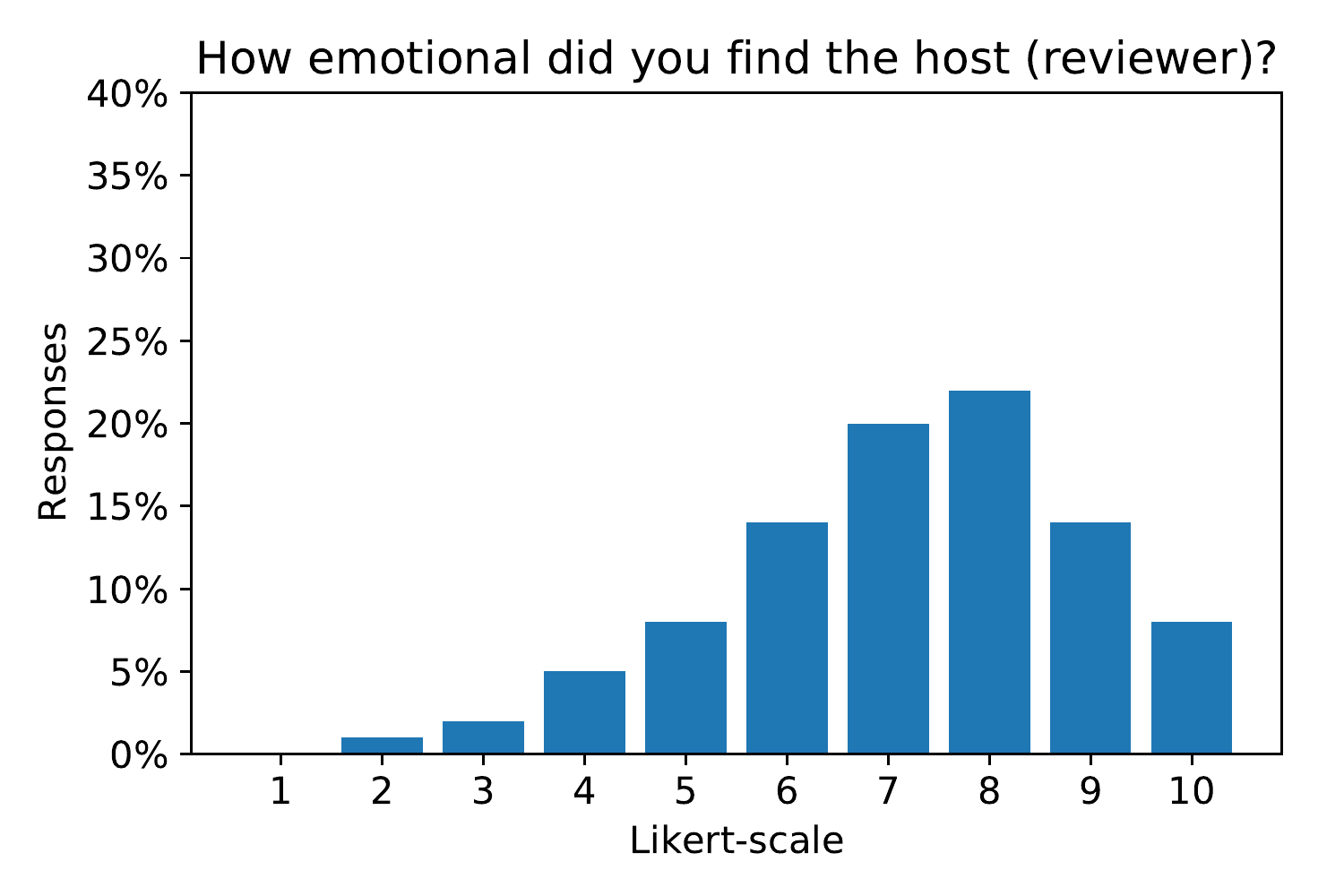} \\
  \includegraphics[width=0.49\linewidth]{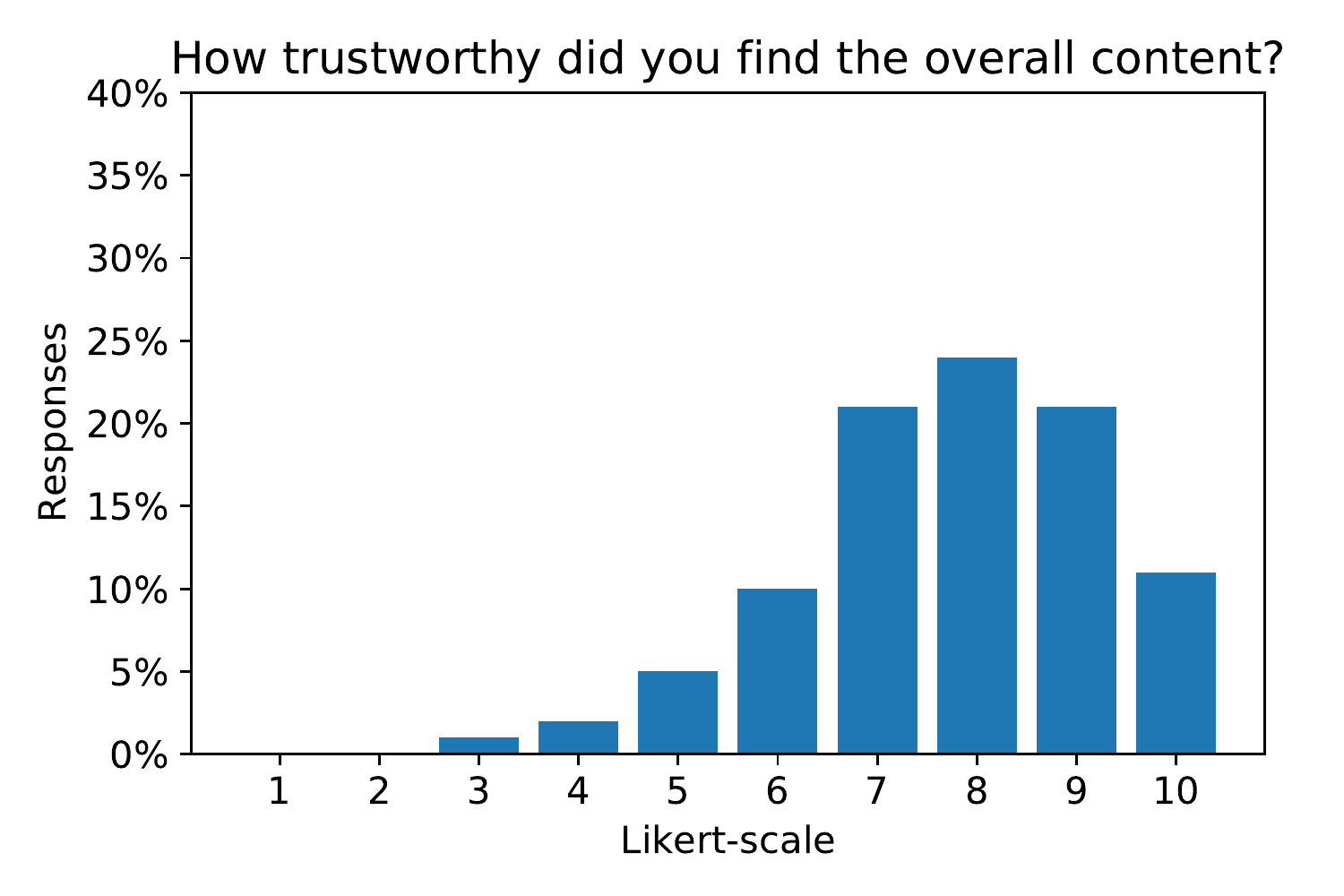}
   \includegraphics[width=0.49\linewidth]{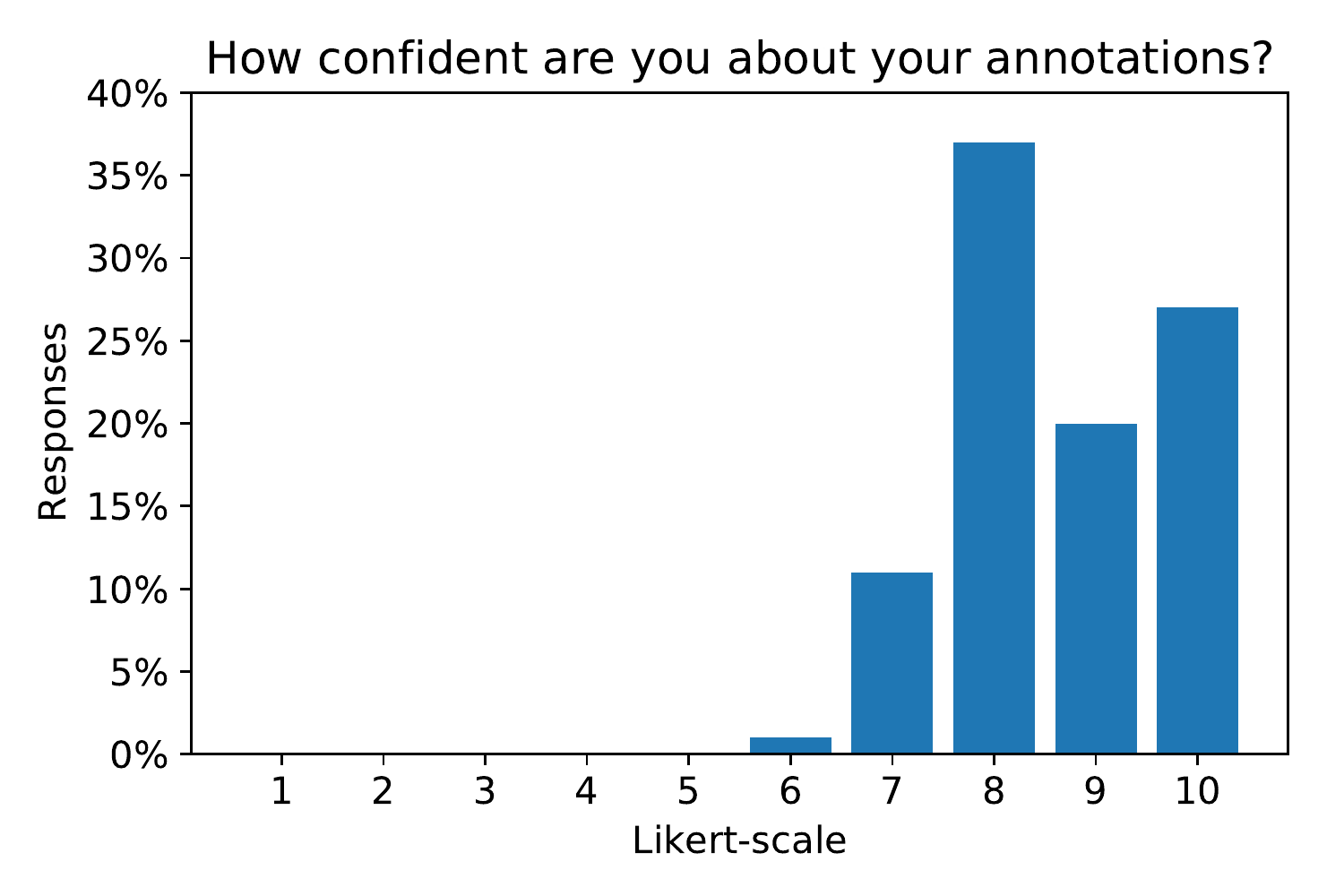}
 \vspace{-0.4cm}
 \caption{Annotation related-metadata of almost 5\,000 ratings, where the annotator gave scores from 1 to 10 regarding the video watched.}
 \vspace{-0.4cm}
 \label{fig:const}
\end{figure}

This data is directly linked to each annotation. An overview of the answers collected is depicted in \Cref{fig:const}. Although being somewhat subjective attributes to evaluate, we see a general agreement across our annotations for all questions with a coefficient of variation between 0.12 and 0.25. 
The appeal of the videos to the annotators seems rather strong with $>$ 20\,\% of the ratings between seven and eight. Similarly, the level of emotionality and trustworthiness portrayed by the hosts were perceived high. Regarding the self-assessment of the annotation quality, the annotators seem very confident in their performance.

\vspace{-1em}
\subsection{First proposed tasks and data availability}
For the MuSe 2020 challenge, we proposed a selection of tasks, and a detailed description can be found in~\cite{stappen2020muse}. To make the (pre-processed) data for each task (\eg specific features) easily accessible after the challenge, we moved them to several Zenodo repositories\footnote{Metadata, and raw material: https://zenodo.org/record/4651164}, a high-speed research data host with storage in the CERN data centre. The \ac{MuSe-CaR} database is available online for researchers who fulfil the requirements of the EULA (\eg academic-use only). 
The specific links can be found in each of the following sub-sections. 


\textbf{Multimodal Sentiment in-the-Wild (\musewild)}\footnote{\musewild: https://zenodo.org/record/4134609}: 
\musewild aims to predict the level of the affective dimensions of Arousal and Valence in a time-continuous manner. Timestamps to enable modality alignment and fusion on word-, sentence-, and utterance-level as well as several acoustic, visual and textual-based features are pre-computed and provided with the task package.

\textbf{Multimodal Emotion-Target Engagement (\musetopic)}\footnote{\musetopic: https://zenodo.org/record/4134733}:
The \emph{\musetopic task} focuses on the prediction of 10-classes of domain-specific (automotive, as given by the chosen database) topics as the target of categorical Valence and Arousal emotions. The three classes (low, medium, and high) of Valence and Arousal are each predicted for every topic segment. These classes are created by averaging the mean value of the temporally aggregated continuous labels of \musewild and then dividing them into three equally sized classes (33\,\%) for each label.

\textbf{Multimodal Trustworthiness (\musetrust)}\footnote{\musetrust): https://zenodo.org/record/4134758}: 
The last task aims to develop methods to predict a continuous Trustworthiness signal in a sequential manner. Aligned Valence and Arousal annotations are also provided, to explore the relationship between all three dimensions \eg by training multi-task networks.

The evaluation metric of \musewild and \musetrust is the CCC, which is often used in similar tasks~\cite{valstar2013avec, ringeval2017avec} as it stands for a theoretically well understood measure\cite{pandit2019many}, able to reflect the reproducibility and performance while being robust to changes in scale and location~\cite{lawrence1989concordance}. \musetopic is measured in a combination of F1 and Unweighted Average Recall. 

\Cref{tab:paritioning} shows the size of the training, validation, and test partitions which consider emotional ratings, speaker/ channel independence, and duration, and come with the packages.  In other words, the same host/speaker does not appear across partitions. Before partitioning, we removed less informative data in a pre-processing step. \musewild and \musetrust only includes parts where an active voice or a visible face are included. Only \musetrust includes non-product related video segments, such as, advertisements which might have an impact on the Trustworthiness perception. \musetopic, the more NLP-related task only includes parts with a speaker topic annotation and transcription.

\begin{table}[t!]
\footnotesize
\vspace{-0.4cm}
  \caption{
Partitioning of the three tasks. Reported are the number of unique videos, and the duration for each task in hh:mm:ss. }
  \begin{tabular}{l|rrrr}
    \toprule
    Partition & No. & \musewild & \musetopic  & \musetrust \\
    \midrule
    Train   & 166 & 22\,:16\,:43 &22\,:35\,:55 & 22\,:45\,:52 \\
    Devel.  &  62 & 06\,:48\,:58 &06\,:49\,:46 & 06\,:52\,:22 \\
    Test    &  64 & 06\,:02\,:20 &06\.:14\,:08 & 06\,:12\,:53 \\
    \hline
    $\sum$    & 291 & 35\,:08\,:01 & 35\,:39\,:49 & 35\,:51\,:07\\
  \bottomrule
\end{tabular}
\label{tab:paritioning}
\vspace{-0.3cm}
\end{table}

\vspace{-1em}
\section{Surpassing the MuSe-Trust baseline}

Although the organisers of MuSe 2020 received several prediction submissions for the MuSe-Trust task, none could surpass the results of the baseline models, resulting in no papers for this task being accepted for the official challenge workshop~\cite{stappen2020summary}. 
In this section, we show that a simple but efficient neural network architecture called \textsc{DeepTrust} utilising a Multi-Attention Head Layer (MAHL) for encoding in addition to a bi-directional (bi) Long Short-Term Memory Recurrent Neural Network (LSTM) with augmentation is suitable for modelling Trustworthiness. We use provided feature sets from the challenge as well as extracting new ones, shown to be effective in the other MuSe tasks, resulting in two different (for acoustic and vision: one handcrafted, and one based on deep representations) feature sets for each modality. We run extensive experiments regarding augmentation, complexity, architecture, and learning the impulse (loss) of such a network. Furthermore, we evaluated the performance of unimodal features and multimodal fusion, as well as the training style (single- vs multi-task learning) for the prediction of Trustworthiness.

\vspace{-1em}
\subsection{Features}

\textbf{Acoustic:} For the handcrafted acoustic feature set, we use the extended Geneva Minimalist Acoustic Parameters Set (\egm)~\cite{eyben2015geneva} provided by the MuSe challenge. It is based on 23 acoustic spectral, cepstral, and prosodic low-level descriptors (LLDs) from which statistical functions are calculated. We extract this 88-dimensional feature vector with a window size of 5 seconds and a hop size of 250\,ms to enable an alignment to the annotation sampling rate. We further apply standardisation to the vector dimensions.

In addition, we extract \vgg functions~\cite{hershey2017cnn} pre-trained on an extensive YouTube audio dataset (AudioSet)\cite{gemmeke2017audio}. The underlying data contains 600 classes and the recordings contain a variety of `in-the-wild' noises that we expect to be beneficial to obtain robust features from our `in-the-wild' videos. By aligning the frame and hope size to the annotation sample rate, we extract a 128-dimensional \vgg embedding vector every 0.25\,s from the underlying log spectrograms.

\textbf{Vision: }The Facial Action Units (FAU), are widely adopted in tasks close to emotion recognition, describing visually perceptible facial movements. FAUs break down facial expressions into 17 individual components of muscle movement, which we obtain from the OpenFace Toolkit~\cite{Baltruvsaitis16-OAO}. For the deep features, the pre-trained \vggf features~\cite{parkhi2015deep}, which were originally developed to identify faces of people are utilised. By removing the last softmax activation layer, we acquire a vector feature representation of the face.

\textbf{Text: }A standard method to transfer words from a symbolic to a continuous representation are word embeddings. These calculate a static, numerical vector per word, depending on the semantics in which the word occurs during training. We extract a 300 dimensional \ft vector for each word of the automatic transcription~\cite{bojanowski2017enriching}. Instead of a static vector representation, context-based NLP transformers extract one vector per word in direct dependence on the context -- during the time of inference. For this technique, we apply a \bert Model~\cite{radford2019language}, well established for a number of NLP tasks, and extract the sum of the last four layers as a 768 dimensional feature vector similar to~\cite{sun2020multi}.

All features are aligned using the timestamps introduced in \Cref{sec:tiers}. 

\vspace{-.5em}
\subsection{\textsc{DeepTrust}: Multihead attention network for Trustworthiness prediction} 

We utilise two neural network mechanisms to model the short- and long-term dependencies of continuous Trustworthiness: enhancing the encoding of the input state by a MHAL and modelling the temporal dynamics of state changes by a LSTM. The attention heads improve the local representation of the extracted features and are able to sustain the long-term (global) dynamics of a sequence in its representation. However, this does not inherit a deeper positional understanding by nature\cite{vaswani2017attention}. To this end, we use the functionality of LSTMs, which are particularly capable of learning short- and mid-term patterns. Similar architectures were used by~\cite{sun2020multi} and~\cite{huang2019efficient} for emotion recognition.

Mathematically, we train a function $F(X_i)$, where $X$ is a sequence of uni- or multimodal input features, which predicts a sequence of regression point estimates $y_i$. For this, we apply $h$ multi-attention heads to obtain more meaningful sequence representations $s_t$, with $T$ being the maximum number of steps. Multi-attention heads are the key building block of transformer networks. In this process, $h$ softmax dot-product attentions (self-attention mechanism, $Att$) are calculated in parallel to learn different discriminative patterns in each head from the three linear projection inputs (query $Q$, key $K$, and values $V$), whereby the division by $\sqrt{d_{k}}$ prevents very small gradients. After scaling, the results of the individual heads are concatenated and fed into a subsequent linear layer $W^{S}$: 
\begin{equation}
\text{MultiHead} = s_t =\text{Concat}\left(\text {head}_{1}, \ldots, \text {head}_{h}\right) W^{S},
\end{equation}

\begin{equation}
\text { head}_{h}=\alpha\left(Q W_{h}^{Q}, K W_{h}^{K}, V W_{h}^{V}\right),
\end{equation}

\begin{equation}
\alpha\text {(Q, K, V)}=\operatorname{softmax}\left(Q K^{T} / \sqrt{d_{k}}\right) V.
\end{equation}

The resulting enhanced sequence $S$ is the input of, \eg a one-directional LSTM
to receive the temporal encoded sequence $O$: 
\begin{equation} \label{tl}
o_t = \overrightarrow{LSTM}(s_t), t \in \{1,...,T\}.
\end{equation}
Finally, the temporally encoded information are feed into a regression layer, predicting $y_i$.

\subsection{Experimental settings}
Unlike most large emotional datasets with continuous annotations, MuSe-CaR does not break the videos down into artificial, equally-sized segments, so that content and context can remain largely intertwined. However, this has the disadvantage that some sequences end up very long (> 5\,000 steps) increasing the amount of computation power needed. To solve this issue and increase the amount of data, the length is segmented to a fixed number of sequence steps $ws$ moving with a hop size $hs$ as proposed by the baseline paper \cite{stappen2020muse} and several participants~\cite{sun2020multi}. Furthermore, we utilise the segment id, as it provides the models with an additional positional encoding~\cite{sun2020multi}.

The models are trained for a maximum of 100 epochs using an Adam optimiser, while the learning rate is reduced, as it reaches a plateau for more than 10 epochs. As in the challenge, the CCC is evaluated on the development set after each epoch, and after the training, the best configurations is subsequently evaluated on the test set. 

As we aim for an integrated approach of modality fusion, we use early (concatenate the inputs) as well as late (using a LSTM) fusion in order to better learn from the interaction of the modalities.

For the ablation study, we choose to set $ws = 200$, $hs = 100$ after initial experiments, and the number of hidden neurons of the a bidirectional LSTM to $64$. If not otherwise stated, we run a hyperparameter optimisation using $h \in \{2,4,8\}$ heads, $lr \in \{0,0001, 0.001, 0.005\}$, a batch size $\in \{512, 1024, 2048\}$ and report the best result indicating the hyperparameter setting by $HP$.

\vspace{-1em}
\subsection{Results}

\textbf{Unimodal: }
First, we compare the performance of our feature sets. As shown in \Cref{tab:uni}, the advanced \bert features and the deep acoustic features \vgg yield higher results in terms of CCC. Only the results on the vision features behave contrary, where the deep face features \vggf seem to easily overfit on the development set, while \au scores are lower, but generalise better. Early fusion of the both text modalities adds a small advantage compared to single use. For all other modalities, the results are worse on the test set. When comparing the model with \egm and \ft features with those of the baseline, both show large improvements on the development set and the result for \ft almost doubled on the test set. For the following experiments, we use the best performing feature set of each modality, namely \bert, \vgg, and \au. 


\begin{table}
 \caption{Results of MuSe-Trust using the devel(opment) and test set,  applying early and late fusion. Results are reported in concordance correlation coefficient (CCC). As feature sets for text: \ft (FT) and \bert; acoustic: \vgg and \egm; vision: \vggf and \au for our models.
 }
 \centering
  \resizebox{0.9\columnwidth}{!}{
 \centering
 \settowidth\rotheadsize{\textbf{\hphantom{sp}Vision}}
 \begin{tabular}{llrrc} 
 \toprule
 & \textbf{Feature sets} & \textbf{Dev.} & \textbf{Test} & \textbf{HP}\\
 \midrule
 \multirow{3}{*}{\rothead{\textbf{\hphantom{spac}Text}}}  
   & \ft & $.4559$ & $.4782$ & I \\
   & \bert & $.5624$ & \textbf{.5539} & I \\ 
   & \bert + \ft & \textbf{.5648} & $.5478$ & II \\ 
  \hline
 \multirow{3}{*}{\rothead{\textbf{\hphantom{sp}Audio}}} 
   & \egm & $.3921$ & $.1220$ & III \\ 
   & \vgg & \textbf{.5376} & \textbf{.4035} & I \\
   & \vgg + \egm & $.4751$ & $.2402$ & I \\ 
   \hline
 \multirow{3}{*}{\rothead{\textbf{\hphantom{sp}Vision}}} 
   & \au & $.3675$ & \textbf{.3623} & I \\ 
   & \vggf & \textbf{.4000} & $.2802$ & IV \\
   & \vggf + \au & $.3936$ & $.3298$ & V \\ 
   \hline
 \bottomrule
 \end{tabular}
 }
 \label{tab:uni}
  \vspace{-.5em}
\end{table}

\textbf{Augmentation: } 
Previous challenges~\cite{valstar2013avec} identified the effective use of the available data as a key performance driver. \Cref{tab:seqlength} shows the results under changing number of sequence steps $ws$ and hop sizes $hs$. When $ws$ is equal $hs$, the sequences have no overlap. A larger $ws$ appears to be generally valuable. Most likely, this variable ($ws$) improves the ability to capture global dynamic changes, leading to a more expressive representation of the state of trust. Having no overlap yields stable, generalisable results, while, when applying an overlap, the results could be either better or worse. 
As a rule of thumb, the longer the sequences are, the higher the overlap can be, while a good estimate ranges between $hs$ = 0.3 -- 0.5 $ws$. However, if $ws$ is small (\eg 100, 200), the results might not be generalisable to test. This might be counteracted by applying additional augmentation to the reappearing sequence steps.


\begin{table}[ht]
 \caption{Results of MuSe-Trust using the devel(opment) and test set. Results are reported in CCC.}
 \centering
  \label{tab:seqlength}
 \resizebox{\columnwidth}{!}{
 \begin{tabular}{cc||rr|rr|rr|rr} 
 \toprule
 \multicolumn{2}{c||}{\textbf{steps}} & \multicolumn{2}{c|}{\textbf{\bert}} & \multicolumn{2}{c|}{\textbf{\vgg}} & \multicolumn{2}{c|}{\textbf{\au}} & \multicolumn{2}{c}{\textbf{\O}} \\
 \textbf{$ws$} & \textbf{$hs$} & \textbf{Dev.} & \textbf{Test} & \textbf{Dev.} & \textbf{Test} & \textbf{Dev.} & \textbf{Test} & \textbf{Dev.} & \textbf{Test} \\
  \midrule
    750&750&.5641&.5540&.4274&.4344&.3775&.3719&.4563&.4534 \\
    750&500&.5739&\textbf{.5747}&.5604&\textbf{.4699}&.3671&.3705&.5005&.4717 \\ 
    750&250&.5889&.5693&.5386&.4686&.4305&\textbf{.4843}&.5193&\textbf{.5074} \\ 
    \hline
    200&200&.5512&.5245&.5566&.4752&.3614&.2710&.4897&.4236 \\
    200&150&.5500&.5533&.5517&.3034&.3558&.3508&.4858&.4025 \\
    200&100&.5624&.5539&.5376&.4035&.3675&.3623&.4892&\textbf{.4399}\\ 
    200&50&.5282&.5160&.5440&.4081&.3319&.1820&.4680&.3687\\ 
    \hline
    100&100&.5167&.5128&.5064&.4445&.3711&.3709&.4647&\textbf{.4427}\\
    100&50&.5312&.5068&.5369&.2918&.3816&.3294&.4832&.3760\\
    100&25&.5233&.5216&.5264&.2517&.3642&.2911&.4713&.3548\\

  \bottomrule
 \end{tabular}
 }
\end{table}

Other side effects are that, as the length of the sequence increases, both the memory requirement and the training time grow. With our standard architecture, a maximum of $ws$ = $750$ is supported on a 32\,GB Memory GPU, which makes it necessary to trade-off performance for usability.

\textbf{Heads: }
In a unimodal setting, the most suitable number of heads varies from modality to modality, with no clear tendency (\eg number of dimensions). The best results (\cf \Cref{tab:head}) are obtained with 4 heads for \bert ($.5495$ on test), 16 heads for \vgg ($.4592$ on test), and 2 heads for \au ($.3774$ on test). On average, 2, 4, and 16 heads perform very similarly on the development set with a slight advantage for 16 heads ($.4352$) and 8 heads ($.4388$) on test data.

\begin{table}[h]
\vspace{-.5em}
 \caption{Results of MuSe-Trust using the devel(opment) and test set. Results are reported in CCC.}
 \centering
  \label{tab:head}
 \resizebox{\columnwidth}{!}{
 \begin{tabular}{c||rr|rr|rr|rr} 
 \toprule
 \multicolumn{1}{c||}{\textbf{heads}} & \multicolumn{2}{c|}{\textbf{\bert}} & \multicolumn{2}{c|}{\textbf{\vgg}} & \multicolumn{2}{c|}{\textbf{\au}} & \multicolumn{2}{c}{\textbf{\O}} \\
 \textbf{
 } & \textbf{Dev.} & \textbf{Test} & \textbf{Dev.} & \textbf{Test} & \textbf{Dev.} & \textbf{Test} & \textbf{Dev.} & \textbf{Test} \\
  \midrule
    \hline
    2& .5698&.4745&.5375&.4368&.3561&\textbf{.3774}&.4878&.4296 \\
    4& .5624&\textbf{.5539}&.5376&.4035&.3675&.3591&.4892&\textbf{.4388} \\
    8& .5539&.5454&.4035&.2671&.3623&.3280& .4399&.3802 \\
    16&.5693&.5112&.5619&\textbf{.4592}&.3548&.3352&.4953&.4352 \\

  \bottomrule
 \end{tabular}
 }
\end{table}

\textbf{Loss:}
Since the loss and the metric are the same (CCC), we also report the Pearson Correlation Coefficient (PCC) and the Root Mean Square Error (RMSE) for this experiment \cf \Cref{tab:loss}. The CCC loss clearly performs better than the MSE and L1 loss, for the \bert and \vgg features as well as for the average results of the two correlation based metrics (CCC and PCC). However, this is not the case for \au where L1 and MSE perform equally or outperform the CCC loss. For RMSE, FAU has such a strong impact that also the average RMSE of both other loss functions are better.

\begin{table}[h]
\vspace{-.5em}
 \caption{Results of MuSe-Trust using the devel(opment) and test set. Results are reported in CCC, PCC, and RMSE.}
 \settowidth\rotheadsize{mse}
 \centering
  \label{tab:loss}
 \resizebox{\columnwidth}{!}{
 \begin{tabular}{cc||rr|rr|rr|rr} 
 \toprule
 \textbf{loss} &
 \textbf{metric} &
 \multicolumn{2}{c|}{\textbf{\bert}} & \multicolumn{2}{c|}{\textbf{\vgg}} & \multicolumn{2}{c|}{\textbf{\au}} & \multicolumn{2}{c}{\textbf{\O}} \\
 \textbf{ } & & \textbf{Dev.} & \textbf{Test} & \textbf{Dev.} & \textbf{Test} & \textbf{Dev.} & \textbf{Test} & \textbf{Dev.} & \textbf{Test} \\
  \midrule
    \hline
    \multirow{3}{*}{\rothead{\textbf{CCC}}} 
        & CCC &.5624&.5539&.5376&.4035&.3675&.3623&\textbf{.4892}&\textbf{.4399}\\
        & PCC &.5684&.5998&.5384&.4421&.3770&.4301&\textbf{.4946}&\textbf{.4907}\\
        & RMSE &.3652&.3485&.3867&.4199&.4693&.4780&.4071&.4155\\       
        \hline

    \multirow{3}{*}{\rothead{\textbf{L1}}} 
        & CCC &.5076&.5211&.3650&.2408&.3678&.3407&.4135&.3675\\
        & PCC&.5432&.5712&.3877&.3270&.3724&.3728&.4344&.4237\\
        & RMSE&.3595&.3409&.4031&.3978&.4350&.3690&.3992&\textbf{.3692}\\
        \hline
    \multirow{3}{*}{\rothead{\textbf{MSE}}}  
        & CCC &.5215&.5433&.3932&.4094&.3537&.3243&.4228&.4257\\
        & PCC &.5455&.5570&.3932&.4100&.3584&.3498&.4324&.4392\\
        & RMSE &.3584&.3470&.3932&.4160&.4407&.3796&\textbf{.3974}&.3809\\    

  \hline
  
  \bottomrule
 \end{tabular}
 }
\end{table}

\textbf{Model:}
Next, we compare several architectural choices on the unimodal feature selections. As we can see in \Cref{tab:table3}, using the combination of both modules is sensible. One configuration (2 MHAL+Bi-LSTM) yields a better result on the \bert test set. For all others and on average, the one-layer MHAL and a bidirectional LSTM architecture achieves the best results.

\begin{table}[h]
 \caption{Results of MuSe-Trust using the devel(opment) and test set. Results are reported in CCC.}
 \centering
  \label{tab:table3}
 \resizebox{\columnwidth}{!}{
 \begin{tabular}{l||rr|rr|rr|rr} 
 \toprule
 \multicolumn{1}{c||}{\textbf{network}} & \multicolumn{2}{c|}{\textbf{\bert}} & \multicolumn{2}{c|}{\textbf{\vgg}} & \multicolumn{2}{c|}{\textbf{\au}} & \multicolumn{2}{c}{\textbf{\O}} \\
 \textbf{
 } & \textbf{Dev.} & \textbf{Test} & \textbf{Dev.} & \textbf{Test} & \textbf{Dev.} & \textbf{Test} & \textbf{Dev.} & \textbf{Test} \\
  \midrule
    \hline
        MHAL &.3117&.3248&.4230& .3150& .3677&.3351&.3675&.3250 \\
        LSTM &.5165&.5170& .5441&.3771&.3270& .2513&.4625&.3818 \\
        MHAL+LSTM & .5423&.5526&.5368&.2248&.3609&.3047&.4800&.3607 \\
        MHAL+2 Bi-LSTM &.5456&.5504&.5259&.3688&.3642&.2973&.4786&.4055 \\
        MHAL+Bi-LSTM &.5624&.5539&.5376&\textbf{.4035}&.3675&\textbf{.3623}&.4892&\textbf{.4399} \\
        2 MHAL+Bi-LSTM &.5548&\textbf{.5762}&.4918&.3818&.3645&.3447&.4704&.4342 \\
        2 MHAL+2 Bi-LSTM &.5410& .5344&.4942&.3233&.3553&.3523&.4635&.4033 \\
        3 MHAL+3 Bi-LSTM &.5437&.5089&.4977&.3376&.3455&.3104&.4623&.3856 \\
  \bottomrule
 \end{tabular}
 }
\end{table}

\textbf{Multimodal fusion:}
Fusing the best single modalities improves all results on the development set. However, the best -- the fusion of \bert and \vgg -- generalises poorly (devel: $.5833$ CCC vs. test: $.5108$ CCC) achieving a lower result than \bert only. In comparison, the fusion of text and vision features generalises well and achieves $.5880$ CCC on test data ($.5803$ CCC on development) -- a better result than every single modality. Similarly, \vgg and \au achieve slightly higher results on both sets when being fused resulting in .4287 CCC on the test set.

\textbf{Multi-task learning:}
Using our approach to predict Arousal, Valence, and Trustworthiness simultaneously, outperforms the baseline by more than $.2$ on development and almost $.15$ on the test set. Adding more weight to the Trustworthiness predictions (II.) improves the result slightly. 

\begin{table}[h]
 \caption{Results of MuSe-Trust using the devel(opment) and test set. Results of Trustworthiness (T), Arousal (A), and Valence (V) reported in CCC. Configurations: (I.) equal loss weight or 0.33 (II.) 0.5 x Trustworthiness, 0.25 * \{Arousal,Valence\}.}
 \centering
  \label{tab:multitask}
 \resizebox{\columnwidth}{!}{
 \begin{tabular}{ll||rr|c|c} 
 \toprule
 \multicolumn{2}{c||}{\textbf{Configuration}} & \multicolumn{2}{c|}{\textbf{T}} & \textbf{A} & \textbf{V} \\
 \textbf{Model} & \textbf{Features} & \textbf{Dev.} & \textbf{Test} & \textbf{Dev.} & \textbf{Dev.} \\
  \midrule
    End2You-Multitask~\cite{stappen2020muse}& \ft + \vggf + A &3264 & .4119&--&-- \\
    MHAL+LSTM-Multi (I.)& \bert + \vgg + \au &.5428&.5456&.4102&.4442\\ 
    MHAL+LSTM-Multi (II.)& \bert + \vgg + \au &.5497&.5518&.4132&.4215\\ 
  \bottomrule
 \end{tabular}
 }
\end{table}

\textbf{Best configuration:}
Our best results are shown \Cref{tab:baseline} achieving on development $.6507$ CCC and on test $.6105$ CCC (late) fusing all except the \vgg predictions of the fully trained uni-modal models. The results  represent a major improvement in terms of the baselines established by~\cite{stappen2020muse} more than 50\,\% on the development set. 

\begin{table}
 \caption{Results of MuSe-Trust using the dev(elopment) and test set. Results are reported in concordance correlation coefficient (CCC). As feature sets for text: \ft (FT) and \bert; acoustic: \vgg, \egm, and DeepSpectrum (DS); vision: \vggf, \au,  and 2D landmarks (2D). Furthermore, the baseline models use the raw audios (A) in End2You and several combined vision features (V) in LSTM + Self-Att. 
 }
  \resizebox{\columnwidth}{!}{
 \centering
 \settowidth\rotheadsize{Vision}
 \begin{tabular}{llrr} 
 \toprule
 \textbf{Model} & \textbf{Features} & \textbf{Dev.} & \textbf{Test} \\
   \hline
  MultiFusion~\cite{yangmultimodal} & \ft + DS + 2D & $.3426$ & $.3259$ \\
  \hline
 \multicolumn{4}{c}{\textbf{Official Baselines~\cite{stappen2020muse}}} \\ \hline  
   LSTM+Self-ATT & \egm & $.1576$ & $.1385$ \\
   LSTM+Self-ATT & \ft & $.2278$ & $.2549$ \\
   LSTM+Self-ATT & \ft + \egm & $.2296$ & $.2054$ \\
   LSTM+Self-ATT & \ft + \egm + V & $.1245$ & $.1695$  \\   
   End2You & \ft + \vggf + A & \textbf{.3198} & \textbf{.4128}  \\ 
   \hline
 \multicolumn{4}{c}{\textbf{Ours (fusion)}} \\ 
  \hline  
  \textsc{DeepTrust} best-of (early) & \bert + \vgg + \au & $.6241$ & $.5073$\\ 
  \textsc{DeepTrust} (early) & \bert + \ft +\vgg + \egm + \au & $.5445$ & $.4998$\\    
  \textsc{DeepTrust} best-of (late) & \bert + \vgg + \au & $.6075$ & $.5796$\\ 
  \textsc{DeepTrust} (late) & \bert + \ft +\vgg + \egm + \au & \textbf{.6507} & \textbf{.6105}\\    
   
 \bottomrule
 \end{tabular}
 }
 \label{tab:baseline}
\end{table}
			
\vspace{-1em}
\section{Future work and limitations}\label{sec:futurework}
In the future, we plan to extend the first proposed tasks in several directions by using the additional levels of annotation presented. Of particular interest 
are the possible presence of a connection between the novel dimension of Trustworthiness and Arousal/ Valence and its usability to evaluate user-generated data, which could give insights on the subjective perception of Trustworthiness.

One focus for the MuSe challenge is to bring together communities from differing computational disciplines; mainly, the sentiment analysis community preferring to predict discrete sentiment/ emotion categories coming from an NLP background~\cite{zadeh2018proceedings}, and the audio-visual emotion recognition community predicting continuous-valued Valence and Arousal dimensions of emotion (circumplex model of affect\cite{russell1980_Circumplex}) originated in intelligent audio and visual signal processing, while often disregarding the potential of the textual modality~\cite{valstar2013avec,ringeval2017avec, kollias2020analysing,schuller2018interspeech}. However, both have its advantages, \eg classes of emotions are more intuitive for humans (happy label vs Arousal and Valence scores) and dimensions are more generaliseable and are highly influenced by related, explicitly multimodal learning, techniques~\cite{arevalo2020gated, gomez2020exploring, qiu2020multimodal}. In theory, the \textbf{russell' circumplex model} of emotion allows a \textbf{mapping} (diarisation) from continuous signals to emotion labels, which is often refuted~\cite{hamann2012mapping}. To date, no reliable approaches exist for dynamically modeling
continuous emotional values to classes on a large scale.

The question arises whether a special mapping can be derived for `in-the-wild' environments. This would also help enabling transfer-learning capabilities between the two categories of research and accompanying datasets.

Connected to the previously introduced direction, the continuous annotation signals also have to be \textbf{summarised} time-wise on \eg sentence- or segment-level for the mapping process. Emotions are an intense feeling that is rather short-term and is typically directed at a \textbf{source/topic}. 
We want to explore suitable ways to \textbf{time-aggregated emotion annotations} to reference topics. 
According fusion and aggregation approaches might even rely directly on the five raw annotations, such as in \ac{DTW}~\cite{rabiner1993fundamentals} and \ac{DCTW}~\cite{trigeorgis2016deep} and might incorporate an explicit model of uncertainty expressed by the (dis-)agreement of annotators. Starting with unsupervised approaches, one might extend the dataset to specific classes in the future.
Furthermore, one should explore ways to generally improve the time and cost of intensive continuous annotation. One idea 
is to learn stable representations of annotators behaviours to create additional artificial annotations based on a reduced number of real annotations.

Since multimodal sentiment analysis `in-the-wild' often utilises user-generated data and a variety of `in-the-wild' characteristics, we want to explore those directions in more depth.

With the collected \textbf{YouTube metadata} of user engagement (\eg view count, like and dislike ratio/count, sentimentality of the video comments), we want to investigate if they can be predicted purely relying on (features of) annotated continuous emotions.

The investigation of `in-the-wild' influences and data imbalance is of further interest. Typically, when developing models for emotion recognition, the aim is to work with a dataset that is as balanced and clean as possible with regard to the people and environment characteristics. This avoids bias in the modelling which leads to worse results when predicting data with an unknown distribution, as well having ethical implications in real-world implementation if not counteracted. When collecting data `in-the-wild', however, this is sometimes not possible. For this reason, the community needs to explore the effects as well as develop appropriate counter-strategies of, \eg gender-wise unbalanced training data. Another example is the relevance of \textbf{facial-related features} in the prediction of Valence. While in most lab settings it is rather unusual for participants to wear glasses or sunglasses, this is more often the case for `in-the-wild' data. Previous studies showed that the degree of occlusion of the face is negatively correlated with the performance of face recognition~\cite{atallah2018face} and the \textbf{presence of occlusion}, such as sunglasses or masks, degrades the performance of facial expression recognition systems~\cite{sajjad2018facial}. We have deliberately collected around 30\,\% hosts wearing glasses or sunglasses in order to be able to deepen these studies and also estimate how multimodal approaches are of help to overcome those kinds of challenges. A similar directions is to measure the influence of \textbf{superimposed banners} which might obscure important video content, such as interaction objects. We want to explore ways to detect, \textbf{artificially remove} and replace such occlusion \eg by utilising Generative Adversarial Networks.

\textbf{Limitations.}
As defined by the categories of~\cite{douglas2005multimodal}, collecting meaningful in-the-wild data is always a trade-off of naturalness and lack of control. Although we provide natural, minutes-long context to capture the change in emotions towards multiple topics instead of short dialogues, we see some limitation regarding the range of possible context. In this corpus, we wanted to connect emotions to overarching topics enabling emotion-context interaction; hence, only selected material of one domain (car reviews) was considered which naturally limits the range of topic. This has also been done in other large emotional datasets as in, \eg advertisement in the case of \cite{kossaifi2019sewa}. We see improvements on the other criteria: Primitive descriptors are only common in affective computing but not in multimodal sentiment. We address the gap by providing two dimensions of real continuous emotions and a user-generated content specific one giving advanced flexibility and help to bridge the two communities. The scope of the dataset (\eg number of hosts, modalities), based on the previous results, does seem to be sufficient to generalise personal-independent affect well and the provided linguistic transcriptions are of high value. Focusing on the user-generated area, we kept a high degree of naturalness, only excluding parts not sensitive to affective recognition (\eg no face and no voice) while keeping all others, even very noisy sections (\eg occlusions, selfie-camera, or half faces etc.).

\vspace{-1em}
\section{Summary and Conclusion}
In this paper, we introduced \ac{MuSe-CaR} -- a multimodal sentiment analysis in real-life media dataset. It was collected in user-generated, noisy environments, and consists of around 300 audio-visual and transcript recordings of more than 70 hosts. We described the extensive annotation process in depth covering 11 tiers including dimensional emotions and layers to model the interaction between them and speaker topics and to visual entities. We intentionally  selected videos containing novel and challenging in-the-wild characteristics, including dynamic backgrounds and changing shots as well as angles of the face.

From this multimodal corpus of emotional car reviews, we derived three initial tasks: i) \musewild, where the level of the affective dimensions of Arousal and Valence has to be predicted; ii) \musetopic, where the domain-related conversational topics and three intensity classes of Arousal and Valence have to be predicted; and, iii) \musetrust, where the level of continuous Trustworthiness has to be predicted. These task are publicly available to the research community, representing a testing bed for efforts in automatic analysis of audio-visual behaviour. In addition, we proposed a simple but efficient network \textsc{DeepTrust} using attention-enhanced encoding to tackle the last task, largely outperforming baseline results of \musetrust. We provided exhaustive experiments and showed it has also  multi-task prediction capabilities which is both helpful in advancing this novel task as well as the field of continuous affect estimation. Finally, we introduced some of our future research directions and limitations of the dataset. We hope this dataset is another valuable extension for the research community and another cornerstone in mastering multimodal sentiment analysis.




\ifCLASSOPTIONcompsoc
  \section*{Acknowledgments}
\else
  \section*{Acknowledgment}
\fi

The authors would like to thank the annotators and students at the University of Augsburg and Imperial College London who contributed to this project in various ways.
This work is partially funded by the the DFG’s Reinhart Koselleckproject No.\ 442218748 (AUDI0NOMOUS) as well as the European Union Horizon 2020 research and innovation programme, grant agreement 856879.

\ifCLASSOPTIONcaptionsoff
  \newpage
\fi



%

\bibliographystyle{IEEEtran}
\bibliography{Main}

\vspace{-2.5em}
\begin{IEEEbiography}[{\includegraphics[width=1in,height=1.25in,clip]{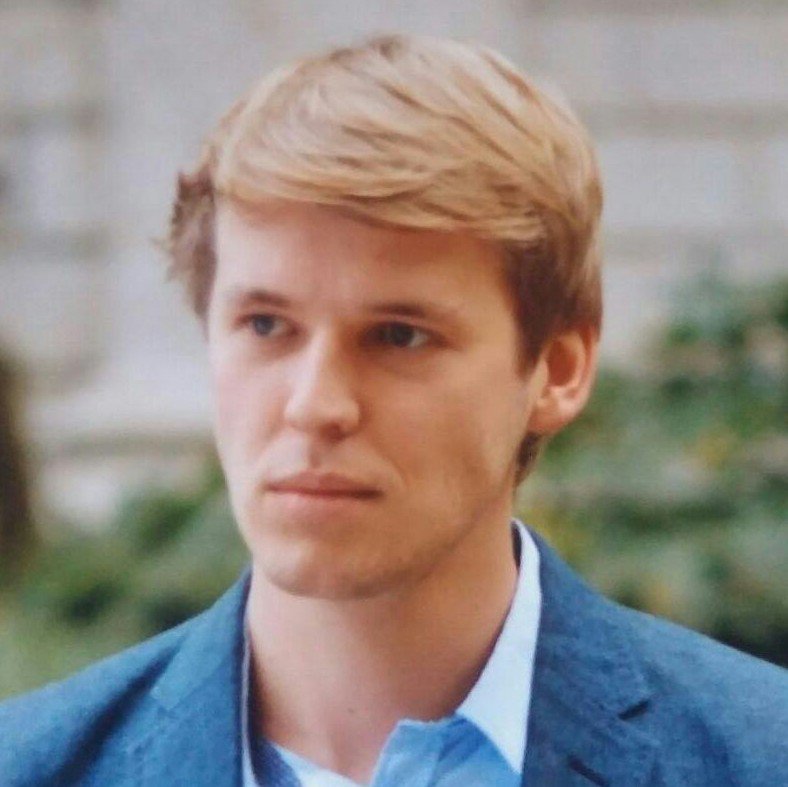}}]{Lukas Stappen}
received his Master of Science in Data Science with distinction from King's College London in 2017. He then joined
the group for Machine Learning in Health Informatics. Currently, he is a PhD candidate at the Chair for Embedded Intelligence for Health Care and Wellbeing, University of Augsburg, Germany, and a PhD Fellow of the BMW Group. His research interests include affective computing, multimodal sentiment analysis, and multimodal/cross-modal representation learning with a core focus 
on `in-the-wild' environments.
\end{IEEEbiography}
\vspace{-2.5em}

\begin{IEEEbiography}[{\includegraphics[width=1in,height=1.25in,clip,keepaspectratio]{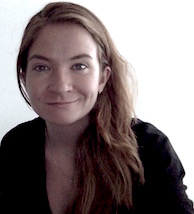}}]{Alice Baird}
received her MFA in Sound Art from Columbia University's Computer Music Center and is currently a Ph.D Fellow of the ZD.B, supervised by Professor Prof.\ Björn Schuller at the Chair of Embedded Intelligence for Healthcare and Wellbeing, University of Augsburg, Germany. Her research is focused on intelligent audio analysis in the domain of speech and general audio, and her research
interests include: health informatics, affective computing, computational paralinguistics, and
speech pathology.
\end{IEEEbiography}
\vspace{-2.5em}
\begin{IEEEbiography}
    [{\includegraphics[width=1in,height=1.25in,clip,keepaspectratio]{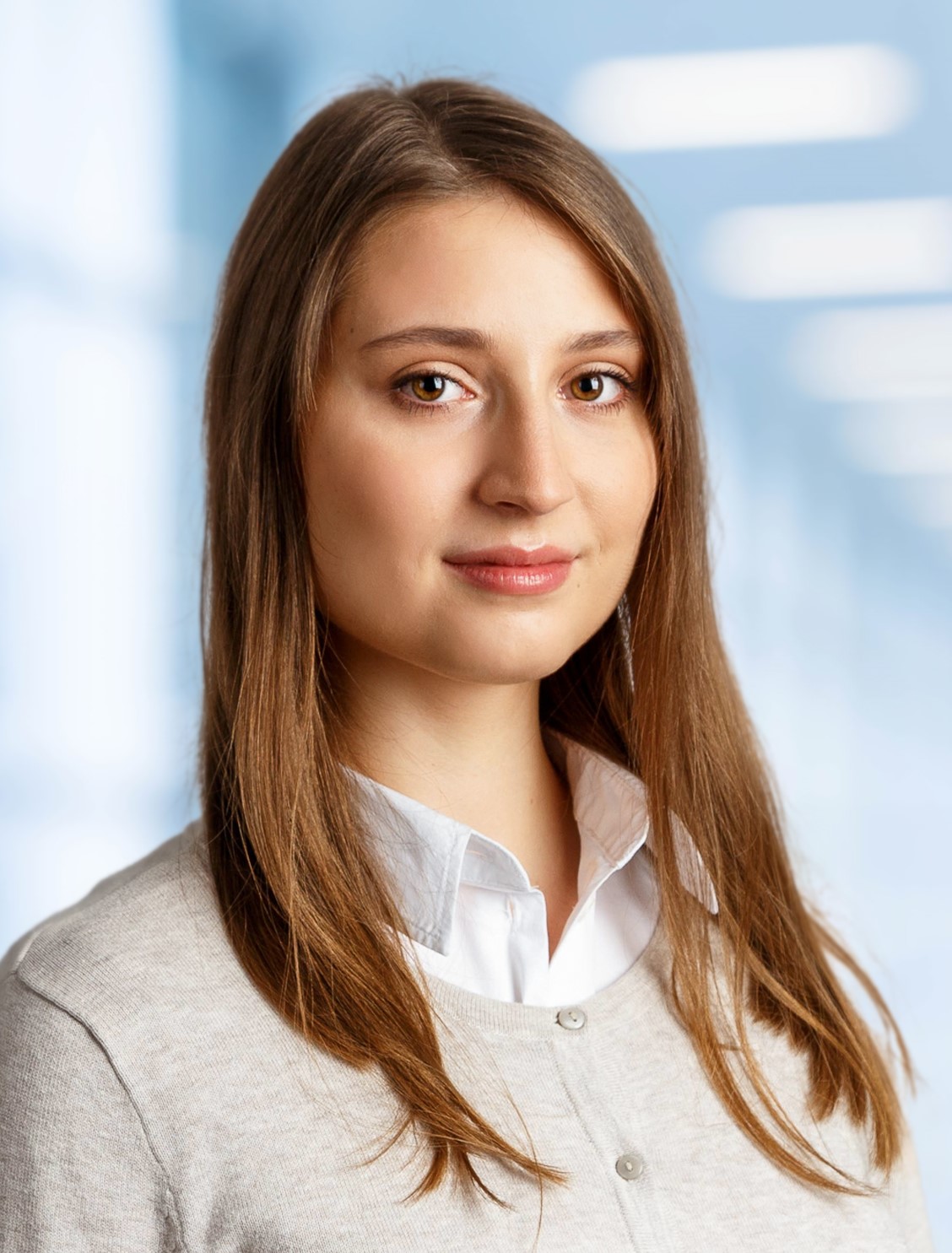}}]{Lea Schumann}
received her B.\,Sc.\ degree in computer science from the University of Augsburg, Germany, in 2018 and is currently working towards her M.\,Sc.\ degree in computer science with a strong focus on deep learning. Her research interests include computer vision, affective computing, and natural language processing.
\end{IEEEbiography}

\vspace{-2.5em}
\begin{IEEEbiography}
    [{\includegraphics[width=1in,height=1.25in,clip,keepaspectratio]{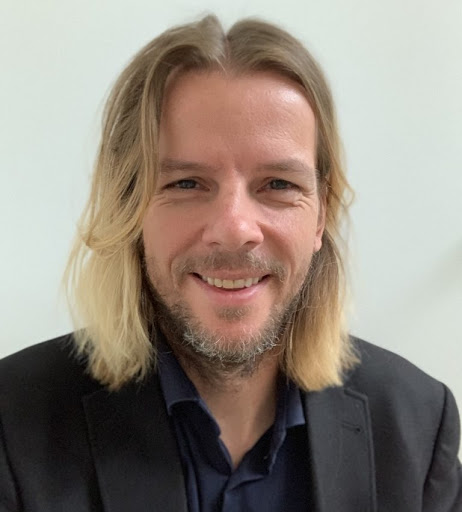}}]{Björn Schuller}
received his diploma, doctoral degree, habilitation, and Adjunct Teaching Professor in Machine Intelligence and Signal Processing all in EE/IT from TUM in Munich/Germany. He is Full Professor of Artificial Intelligence and the Head of GLAM at Imperial College London/UK, Full Professor and Chair of Embedded Intelligence for Health Care and Wellbeing at the University of Augsburg/Germany, and permanent Visiting Professor at HIT/China amongst other Professorships and Affiliations. Previous stays include Full Professor at the University of Passau/Germany, Researcher at Joanneum Research in Graz/Austria, and the CNRS-LIMSI in Orsay/France. He is a Fellow of the IEEE and Golden Core Awardee of the IEEE Computer Society, Fellow of the BCS, Fellow of the ISCA, President-Emeritus of the AAAC, and Senior Member of the ACM. He (co-)authored 1\,000+ publications (35k+ citations, h-index=85), is Field Chief Editor of Frontiers in Digital Health and was Editor in Chief of the IEEE Transactions on affective computing amongst manifold further commitments and service to the community. His 30+ awards include having been honoured as one of 40 extraordinary scientists under the age of 40 by the WEF in 2015. 
\end{IEEEbiography}

%








\end{document}